\documentclass[twocolumn,aps,pra,10pt]{revtex4-2}
\usepackage{amssymb}
\usepackage{amsfonts}
\usepackage{amsmath}
\usepackage{amsxtra}
\usepackage{amscd}
\usepackage{amsthm}
\usepackage{graphicx}
\usepackage{epsf}
\usepackage{bbold}
\usepackage{xcolor}
\usepackage{array}

\newcommand{\PreserveBackslash}[1]{\let\temp=\\#1\let\\=\temp}
\newcolumntype{C}[1]{>{\PreserveBackslash\centering}p{#1}}
\newcolumntype{R}[1]{>{\PreserveBackslash\raggedleft}p{#1}}
\newcolumntype{L}[1]{>{\PreserveBackslash\raggedright}p{#1}}

\begin{document}

\title{A non-Hermitian Su-Schrieffer-Heeger model with the energy levels of free parafermions}

\author{Edward McCann}
 \email{ed.mccann@lancaster.ac.uk}
\affiliation{Department of Physics, Lancaster University, Lancaster, LA1 4YB, United Kingdom}

\begin{abstract}
Using a parent Hermitian tight-binding model on a bipartite lattice with chiral symmetry, we theoretically generate non-Hermitian models for free fermions with $p$ orbitals per unit cell satisfying a complex generalization of chiral symmetry. The $p$ complex energy bands in $k$ space are given by a common $k$-dependent real factor, determined by the bands of the parent model, multiplied by the $p$th roots of unity. When the parent model is the Su-Schrieffer-Heeger (SSH) model, the single-particle energy levels are the same as those of free parafermion solutions to Baxter's non-Hermitian clock model. This construction relies on fully unidirectional hopping to create Bloch Hamiltonians with the form of generalized permutation matrices, but we also describe the effect of partial unidirectional hopping. For fully bidirectional hopping, the Bloch Hamiltonians are Hermitian and may be separated into even and odd parity blocks with respect to inversion of the orbitals within the unit cell. Partially unidirectional hopping breaks the inversion symmetry and mixes the even and odd blocks, and the real energy spectrum evolves into a complex one as the degree of unidirectionality increases, with details determined by the topology of the parent model and the number of orbitals per unit cell, $p$. We describe this process in detail for $p=3$ and $p=4$ with the SSH model. We also apply our approach to graphene, and show that $AA$-stacked bilayer graphene evolves into a square root Hamiltonian of monolayer graphene with the introduction of unidirectional hopping. We show that higher-order exceptional points occur at edge states and solitons in the non-Hermitian SSH model, and at the Dirac point of non-Hermitian graphene.
\end{abstract}

\maketitle

\begin{figure}[t]
\includegraphics[scale=0.50]{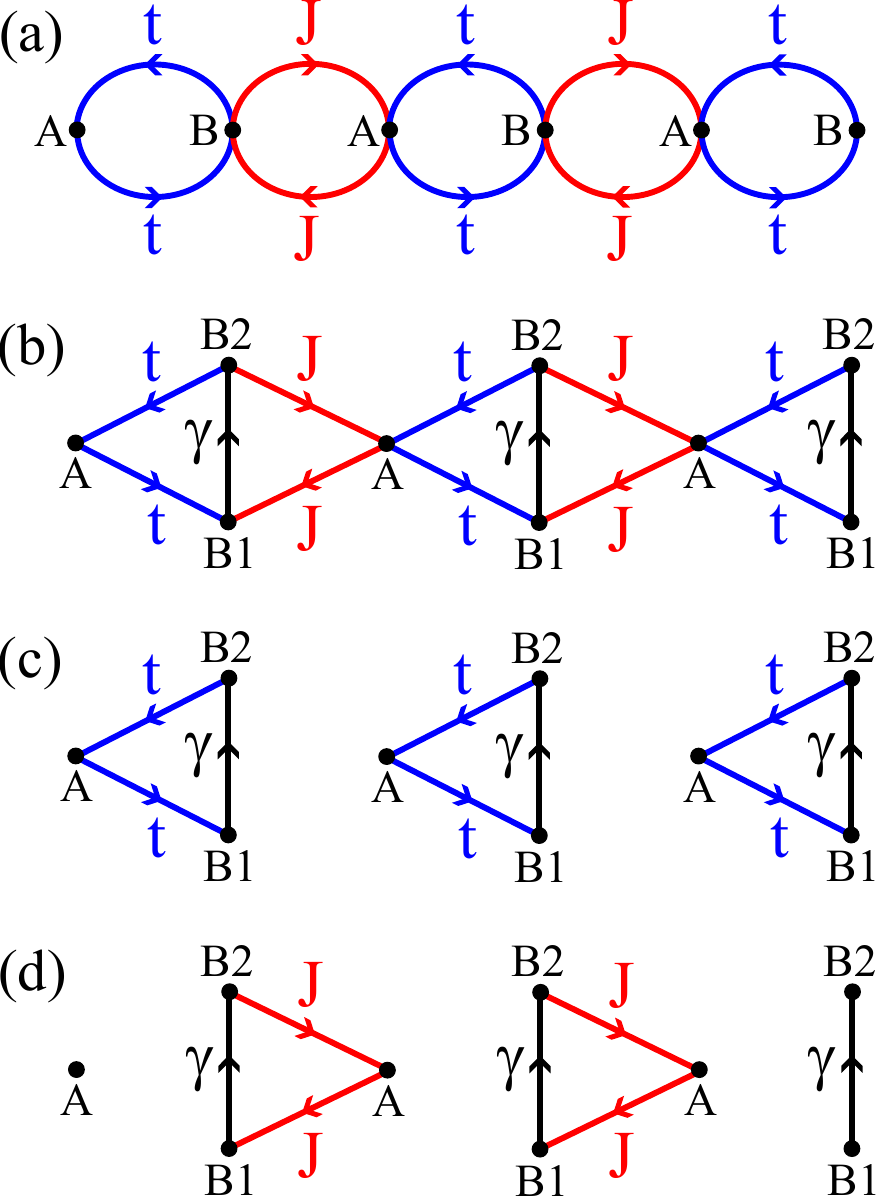}
\caption{(a) Hermitian SSH model with two orbitals per unit cell on sublattices $A$ and $B$ with intracell hopping $t \geq 0$ and intercell hopping $J \geq 0$.
(b) Non-Hermitian Hamiltonian formed from the parent SSH model with two orbitals $B1$ and $B2$ connected by unidirectional hopping $\gamma > 0$ as indicated by the arrows, with unidirectional hopping from $A$ to $B1$ and from $B2$ to $A$.
(c) The topologically trivial phase in the trimer limit with $J = 0$, where each trimer has three states with energies $\epsilon^3 = \gamma t^2$. (d) The topologically nontrivial phase in the trimer limit with $t = 0$, where each trimer has three states with energies $\epsilon^3 = \gamma J^2$, and there are three edge states with energy $\epsilon = 0$.
}\label{nhfig1}
\end{figure}

\section{Introduction}

The Su-Schrieffer-Heeger (SSH) model~\cite{ssh79,ssh80,asboth16} is an Hermitian tight-binding model of noninteracting spinless fermions in one dimension with staggered nearest-neighbor hopping and two orbitals per unit cell. It satisfies time-reversal, charge-conjugation and chiral symmetries, placing it in the BDI class of topological insulators~\cite{schnyder08,kitaev09,ryu10,chiu16}.
Non-Hermitian SSH models have been considered by adding additional tight-binding parameters, usually in one of two ways. The first is to add alternating complex onsite energies which break chiral symmetry but satisfy parity-time (PT) symmetry and may preserve a real energy eigenvalue spectrum~\cite{bender98,hu11,liang13,schomerus13,zhu14,weimann17,klett17,lieu18,halder23,ye24,slootman24}.
The second approach is to add terms which introduce unidirectional hopping~\cite{hatano96,hatano97,hatano98,longhi15b,yao18,song19,liu19b}, an imbalance in the left-moving and right-moving hopping parameters, breaking PT symmetry but preserving chiral symmetry~\cite{lieu18,han21,bergholtz21,halder23,ryu24,ye24}.
Novel phenomena include exceptional points and the non-Hermitian skin effect~\cite{kato66,heiss04,berry04,klaiman08,heiss12,lee16,weimann17,yao18,lieu18,song19,liu19b,kawabata19,miri19,han21,bergholtz21,wiersig22a,wiersig22b,manna23,wiersig23,halder23,sayyad23,ryu24,ye24,schomerus24,bid25a,hetenyi25,kullig25,bid25b,bid25c,gohsrich24}.

In this paper, we consider non-Hermitian Hamiltonians for non-interacting fermions which satisfy a complex generalization of chiral symmetry~\cite{fendley14,marques22,viedma24,gohsrich24} expressed, for a Bloch Hamiltonian ${\cal H} (k)$ in $k$ space, as
\begin{eqnarray}
Z {\cal H} (k) Z^{-1} = \omega {\cal H} (k) , \qquad \omega = \exp (2 \pi i / p) ,
\label{chiralk}
\end{eqnarray}
where $p \geq 2$ is an integer. This chiral symmetry guarantees that the energy spectrum separates into $p$ sectors, and bands in different sectors are related by a common real part multiplied by the $p$th roots of unity.

For simplicity, we consider $p$ orbitals per unit cell so that ${\cal H} (k)$ is a $p \times p$ matrix with $p$ energy bands,
and the chiral operator $Z$ is a $p \times p$ generalization of the diagonal Pauli matrix, $Z = \text{diag} (1 , \omega , \omega^2 , \ldots , \omega^{p-1} )$.
Then, the chiral symmetry~(\ref{chiralk}) may generally be satisfied with ${\cal H} (k)$ in the form of a generalized permutation matrix~\cite{marques22} containing $p$ independent functions of $k$.
Instead of a general Hamiltonian, however, we propose a theoretical scheme to generate the non-Hermitian models from parent Hermitian models on a bipartite lattice.
By introducing unidirectional hopping~\cite{hatano96,hatano97,hatano98,longhi15b} and increasing the number of orbitals to $p$, the $p$ complex energy bands are given by a common $k$-dependent real factor, determined by the bands of the parent model, multiplied by the $p$th roots of unity.

An example where the parent model is the SSH model is illustrated in Fig.~\ref{nhfig1}. Figure~\ref{nhfig1}(a) shows the Hermitian SSH model with two orbitals per unit cell on sublattices $A$ and $B$ with intracell hopping $t \geq 0$ and intercell hopping $J \geq 0$, where the hopping is bidirectional.
The non-Hermitian SSH model with $p=3$ is shown in Fig.~\ref{nhfig1}(b) where site $B$ supports two orbitals, $B1$ and $B2$, connected by unidirectional hopping $\gamma > 0$, with unidirectional hopping from $A$ to $B1$ and from $B2$ to $A$.
The unidirectional hopping is implemented so that it circulates in a counter-clockwise direction in every intracell trimer and in a clockwise direction in every intercell trimer, as if created by an imaginary gauge field~\cite{hatano96,hatano97,hatano98,longhi15b}.
For fully unidirectional hopping, the non-Hermitian model has an energy spectrum related to that of the parent SSH model, and, as with the parent SSH model~\cite{asboth16}, bulk-edge correspondence may be understood pictorially in the trimer limit. For $t \neq 0$, $J=0$, Fig.~\ref{nhfig1}(c), the system splits into separate trimers, each with energy $\epsilon^3 = \gamma t^2$, and there are no edge states (the trivial phase). However, for $t=0$, $J \neq 0$, Fig.~\ref{nhfig1}(d), the system splits into separate trimers, each with energy $\epsilon^3 = \gamma J^2$, plus $p=3$ edge states with energy $\epsilon = 0$ (the nontrivial phase): two at the right edge and one at the left edge.
This is an exceptional point~\cite{kato66,heiss04,berry04,klaiman08,heiss12,weimann17,lieu18,kawabata19,miri19,bergholtz21,wiersig22a,wiersig22b,wiersig23,halder23,sayyad23,viedma24,ryu24,ye24,schomerus24,bid25a,kullig25,bid25b,bid25c}:
The right edge supports a defective eigenvalue at zero energy with an algebraic multiplicity of two, as defined by the roots of the characteristic polynomial, but, owing to the unidirectional coupling of $\gamma$ from $B1$ to $B2$, there is only one linearly-independent eigenvector.
In fact, this is an example of a fragmented exceptional point~\cite{bid25a,bid25b} because the left edge also supports a zero-energy state so, overall, the algebraic multiplicity is three with only two linearly-independent eigenvectors.

It is generally difficult to realize unidirectional hopping in electronic condensed matter systems~\cite{hatano97,hatano98,longhi15b}. 
However, there have been theoretical proposals and experimental realizations of partially unidirectional hopping in other platforms including optical~\cite{longhi15a,longhi15b,weidemann20,qin20,lin21,weidemann22,liu22,gao23,ke23,viedma24}, acoustic~\cite{zhang21a,gao22,gu22}, cold atom~\cite{wu14,li19,liu19a}, and topolectrical~\cite{hofmann19,helbig20,liu21,rafiulislam21,zhang21b,zou21,xie25,sahin25} systems.
The non-Hermitian model with $p=3$ and the SSH model as the parent model, Fig.~\ref{nhfig1}(b), bears some similarity to the model discussed in Ref.~\cite{viedma24} as a cube root of the SSH model, although their model has six sites per unit cell (instead of three) and different values of the unidirectional hopping parameters as compared to $H^{(1,2)}$.
Ref.~\cite{viedma24} proposes the implementation of their model using photonic ring resonators~\cite{longhi15a,longhi15b,qin20,lin21,liu22}, and they estimate that it is possible to achieve very close to perfect unidirectionality.
Motivated by these works on experimental realizations, we consider the influence of partially unidirectional hopping in Sec.~\ref{s:partial}.

Section~\ref{s:full} details the construction of non-Hermitian Hamiltonians with $p$ orbitals and fully unidirectional hopping which satisfy the complex chiral symmetry~(\ref{chiralk}).
We show that it is possible to construct an arbitrary $n$th root model of the parent~\cite{arkinstall17,pelegri19,kremer20,ezawa20,lin21,marques21,viedma24,li24,huang25,zhao25,song25} and we describe the occurrence of exceptional points~\cite{kato66,heiss04,berry04,klaiman08,heiss12,weimann17,lieu18,kawabata19,miri19,bergholtz21,wiersig22a,wiersig22b,wiersig23,halder23,sayyad23,viedma24,ryu24,ye24,schomerus24,bid25a,kullig25,bid25b,bid25c}.
When the parent model is the SSH model, the single-particle energy levels in position space with open boundary conditions are the same as those of free parafermion solutions to Baxter's non-Hermitian clock model~\cite{baxter89a,baxter89b,fendley14}.

Section~\ref{s:partial} describes the role of partially unidirectional hopping
which breaks the complex chiral symmetry~(\ref{chiralk}).
Nevertheless, the complex energy spectrum is constrained by time-reversal symmetry (TRS) and, for an even number of orbitals $p$, by sublattice symmetry.
For fully bidirectional hopping, the constructed model is Hermitian, and it can be block diagonalized into even and odd parity blocks with respect to inversion of the orbitals within the unit cell.
Partially unidirectional hopping breaks the inversion symmetry and mixes the even and odd blocks, and the real energy spectrum evolves into a complex one as the degree of unidirectionality increases.
This process is determined by the topology of the parent model and by the number of orbitals per unit cell, $p$, which we describe in detail for $p=3$ and $p=4$ with the example of the SSH model.
When the SSH model is the parent model, there are real energy levels due to states localized at edges in the topological phase or localized on solitons.

Section~\ref{s:graphene} considers the application to graphene~\cite{novoselov04,castroneto09,mccann13}.
For fully unidirectional hopping, it is possible to construct an arbitrary $n$th root Hamiltonian of graphene with exceptional points at the Dirac points characterized by $2n$ complex bands with dispersion $\epsilon \sim |{\bf q}|^{1/n}$ for small wave vector ${\bf q}$ and Berry's phase~$\pi$~\cite{berry84,liang13,shen18,fan20,tsubota22,longhi23}.
In particular, we show how the introduction of unidirectional hopping changes Hermitian $AA$-stacked bilayer graphene~\cite{lee08,rozhkov16} into a square root Hamiltonian of monolayer graphene.

\section{Fully unidirectional hopping}\label{s:full}

\subsection{General form of the non-Hermitian Bloch Hamiltonian}\label{s:general}

In this section, we describe the construction of the non-Hermitian model when the hopping is fully unidirectional.
We consider a Hamiltonian $H^{(m,n)} (u)$ for a system with $m$ orbitals on the $A$ site, $A1$, $A2$, $\ldots$, $Am$, $n$ orbitals on the $B$ site, $B1$, $B2$, $\ldots$, $Bn$, and $p = m + n$ orbitals per unit cell.
The parameter $0 \leq u \leq 1$ indicates the degree of unidirectionality in the hopping with $u=0$ being the Hermitian limit of bidirectional hopping and $u=1$ being the limit of fully unidirectional hopping.
The Hamiltonian is non-Hermitian for $u >0$.
With translational invariance and periodic boundary conditions, it is possible to Fourier transform the Hamiltonian as $H^{(m,n)} (u) = \sum_k c^{\dagger}_k {\cal H}^{(m,n)} (k,u) c_k$ where $c^{\dagger}_k = \begin{pmatrix} c^{\dagger}_{k,A1} & \ldots & c^{\dagger}_{k,Am} & c^{\dagger}_{k,B1} & \ldots & c^{\dagger}_{k,Bn} \end{pmatrix}$
and the Bloch Hamiltonian ${\cal H}^{(m,n)} (k,u)$ is a $p \times p$ matrix. For the rest of this section, we consider $u=1$.

We consider a parent model which is a Hermitian tight-binding model on a bipartite lattice ($p=2$), with $A$ and $B$ sublattices and a $2 \times 2$ Bloch Hamiltonian of the form
\begin{eqnarray}
{\cal H}^{(1,1)} (k,0) = \begin{pmatrix}
0 & h^{\ast} (k) \\
h (k) & 0
\end{pmatrix} . \label{parent}
\end{eqnarray}
This satisfies chiral symmetry as $\sigma_z {\cal H}^{(1,1)} (k,0) \sigma_z = - {\cal H}^{(1,1)} (k,0)$ where $\sigma_z$ is the diagonal Pauli spin matrix, and it has two energy bands $\epsilon_j^{(1,1)} (k,0) = (-1)^j |h(k)|$ for $j =1,2$.

A non-Hermitian model is formed from the parent by adding additional orbitals on either the $A$ or the $B$ site and introducing unidirectional hopping. For the example of the SSH model with $p=3$, Fig.~\ref{nhfig1}(b), the non-Hermitian model is formed by adding an additional orbital on site $B$, to give orbitals $B1$ and $B2$ on site $B$, with a new Bloch Hamiltonian in the $A$, $B1$, $B2$ basis,
\begin{eqnarray}
{\cal H}^{(1,2)} (k,1) = \begin{pmatrix}
0 & 0 & h^{\ast} (k) \\
h (k) & 0 & 0 \\
0 & \gamma & 0
\end{pmatrix} , \label{h12u1}
\end{eqnarray}
where $\gamma > 0$ is the amplitude for unidirectional hopping from $B1$ to $B2$. This Hamiltonian satisfies a complex generalization of chiral symmetry, $Z {\cal H}^{(1,2)} (k,1) Z^{-1} = \omega {\cal H}^{(1,2)} (k,1)$, where $Z$ is a $3 \times 3$ generalization of $\sigma_z$, $Z = \text{diag} (1 , \omega , \omega^2 )$ with $\omega = \exp (2 \pi i / 3)$~\cite{fendley14,marques22,viedma24}.
There are three complex energy bands related to those of the parent model as $\epsilon_j^{(1,2)} (k,1) = \omega^j (\gamma |h(k)|^2)^{1/3}$ for $j = 1,2,3$, and the topology of the model is related to that of the parent as illustrated in Fig.~\ref{nhfig1}(c) and (d).

In general, the Bloch Hamiltonian ${\cal H}^{(m,n)} (k,1)$ is a $p \times p$ matrix, $p = m+n$, with matrix elements given by
\begin{eqnarray}
{\cal H}_{m+1,m}^{(m,n)} (k,1) = \left( {\cal H}_{1,m+n}^{(m,n)} (k,1) \right)^{\ast} &=&  h(k) , \label{hgen} \\
{\cal H}_{\ell+1,\ell}^{(m,n)} (k,1) &=& \gamma , \nonumber
\end{eqnarray}
for $\ell = 1,2,\ldots , m-1$ and $\ell = m+1,\ldots , p-1$.
All other matrix elements are zero.
This matrix is a generalized permutation matrix with only one non-zero entry in each row and each column. It has an element on the end of the first row, ${\cal H}_{1,m+n}^{(m,n)} (k,1) = h^{\ast} (k)$, and all other non-zero matrix elements are along the lower diagonal.
As a result, the Hamiltonian satisfies the complex generalization of chiral symmetry~(\ref{chiralk}).

The Bloch Hamiltonian ${\cal H}^{(m,n)} (k,1)$ may be viewed as a $p$th root of the parent model squared $\left( {\cal H}^{(1,1)} (k,0) \right)^2$, generalizing the concept of square root Hamiltonians~\cite{arkinstall17,pelegri19,kremer20,ezawa20,lin21,marques21,viedma24,li24,huang25,zhao25,song25}. Chiral symmetry behaves as a unitary symmetry when applied to the Bloch Hamiltonian raised to the $p$th power, $Z \left( {\cal H}^{(m,n)} (k,1) \right)^{p} Z^{-1} = \left( {\cal H}^{(m,n)} (k,1) \right)^{p}$, block diagonalizing it into $p$ separate blocks.
Hence, $\left( {\cal H}^{(m,n)} (k,1) \right)^{p}$ is diagonal~\cite{marques22},
\begin{eqnarray}
\left( {\cal H}^{(m,n)} (k,1) \right)^{p} = \gamma^{p-2} \left| h(k) \right|^2 \mathbb{I}_{p} ,
\end{eqnarray}
where $\mathbb{I}_{p}$ is the $p \times p$ identity matrix.
Thus, $p$ complex energy bands $\epsilon_j^{(m,n)} (k,1)$, $j = 1,2,\ldots , p$, of ${\cal H}^{(m,n)} (k,1)$ are given by the distinct solutions of
\begin{eqnarray}
\left( \epsilon_j^{(m,n)} (k,1) \right)^{p} = \gamma^{p-2} \left| h(k) \right|^2 . \label{pdis2}
\end{eqnarray}

The Bloch Hamiltonian ${\cal H}^{(n,n)} (k,1)$ is an $n$th root of the parent Hamiltonian ${\cal H}^{(1,1)} (k,0)$.
This means that $\left( {\cal H}^{(n,n)} (k,1) \right)^n$ may be written as being block diagonal, using a unitary transformation to re-order the basis, with $n$ separate blocks, each of which is a $2 \times 2$ matrix, $\gamma^{n-1} {\cal H}^{(1,1)} (k,0)$.

\subsection{Exceptional points}\label{s:spectral}

The Bloch Hamiltonian ${\cal H}^{(m,n)} (k,1)$ has an exceptional point at $h(k) = 0$~\cite{kato66,heiss04,berry04,klaiman08,heiss12,weimann17,lieu18,kawabata19,miri19,bergholtz21,wiersig22a,wiersig22b,wiersig23,halder23,sayyad23,viedma24,ryu24,ye24,schomerus24,bid25a,kullig25,bid25b,bid25c}.
In particular, it has defective eigenvalues whereby their algebraic multiplicity, as defined by the roots of the characteristic polynomial, is greater than their geometric multiplicity, the number of associated linearly-independent eigenvectors. For $u=1$ and $h(k) = 0$, all eigenvalues of ${\cal H}^{(m,n)} (k,1)$ are at zero energy with an algebraic multiplicity of $p$ and a geometric multiplicity of two with partial degeneracies~\cite{bid25a,bid25b} $(l_1 , l_2 ) = (m,n)$ associated with the $A$ and $B$ sites, respectively.
For the SSH model, $h(k) \neq 0$ for the bulk in the gapped phases, but the defective eigenvalues may be manifested in position space at edges, as described in Sec.~\ref{s:h12}, and on domain walls, Sec.~\ref{s:soliton}.
For graphene, exceptional points occur at the Dirac point of the parent model (monolayer graphene) in $k$ space where $h(k) = 0$, Sec.~\ref{s:graphene}.

\subsection{The non-Hermitian SSH model}\label{s:ssh}

The Hermitian SSH model has the form of ${\cal H}^{(1,1)} (k,0)$, Eq.~(\ref{parent}), with $h(k) = t + J e^{ika}$ where $t \geq 0$ describes intracell hopping, $J \geq 0$ is intercell hopping, and $a$ is the lattice constant. The lattice in position space is shown in Fig.~\ref{nhfig1}(a).
As well as chiral symmetry~(\ref{chiralk}) with $Z = \sigma_z$ (the diagonal Pauli spin matrix) and $\omega = -1$, the model has time-reversal symmetry $\left( {\cal H}^{(1,1)} (k,0) \right)^{\ast} = {\cal H}^{(1,1)} (-k,0)$ and charge-conjugation symmetry
$\sigma_z \left( {\cal H}^{(1,1)} (k,0) \right)^{\ast} \sigma_z = - {\cal H}^{(1,1)} (-k,0)$.
It has two energy bands given by Eq.~(\ref{pdis2}),
\begin{eqnarray}
\epsilon_j^{(1,1)} (k) = (-1)^j \sqrt{ t^2 + J^2 + 2 t J \cos (ka) }  .
\end{eqnarray}

The non-Hermitian model~(\ref{hgen}) has time-reversal symmetry $\left( {\cal H}^{(m,n)} (k,1) \right)^{\ast} = {\cal H}^{(m,n)} (-k,1)$, generalized chiral symmetry~(\ref{chiralk}) and generalized charge-conjugation symmetry
$Z \left( {\cal H}^{(m,n)} (k,1) \right)^{\ast} Z^{-1} = \omega {\cal H}^{(m,n)} (-k,1)$ where $\omega = \exp (2 \pi i / p)$.
There are $p$ complex energy bands $\epsilon_j^{(m,n)} (k)$, $j = 1,2,\ldots , p$, given by
\begin{eqnarray}
\!\!\!\left( \epsilon_j^{(m,n)} (k) \right)^{p} = \gamma^{p-2} \left( t^2 + J^2 + 2 t J \cos (ka) \right) \! , \label{pdis3}
\end{eqnarray}
which are gapless at $k = \pi / a$ for $t = J$, as for the SSH model.

\subsection{Relation to free parafermions}

When the SSH model is the parent model, the single-particle energy levels are the same as those of free parafermions in Baxter's clock model~\cite{baxter89a,baxter89b,fendley14}. This connection may be understood by considering the form of the Hamiltonian $H^{(1,p-1)} (u=1)$ in position space,
\begin{eqnarray}
H^{(1,p-1)} (u=1) = \sum_{j,m = 1}^{pL} c_j^{\dagger} {\cal H}_{j,m}^{(1,p-1)} c_{m} , \label{hnmatrix}
\end{eqnarray}
where $c_j^{\dagger}$ and $c_j$ are creation and annihilation operators for spinless fermions on site $j$, $p$ is the number of orbitals per unit cell, $L$ is the number of unit cells, and ${\cal H}^{(1,p-1)}$ is a $pL \times pL$ matrix.
Intercell terms are given by
\begin{eqnarray*}
{\cal H}_{(\ell -1) p + 1 ,\ell p}^{(1,p-1)} = {\cal H}_{(\ell -1) p + 2 , (\ell -1) p + 1}^{(1,p-1)} &=& t , \\
{\cal H}_{(\ell - 1)p+m+2,(\ell - 1)p+m+1}^{(1,p-1)} &=& \gamma ,
\end{eqnarray*}
for $\ell = 1,2, \ldots , L$ and $m = 1,2,\ldots , p-2$, and intracell terms are given by
\begin{eqnarray*}
{\cal H}_{(\ell - 1)p+2,\ell p+1}^{(1,p-1)} =
{\cal H}_{\ell p+1,\ell p}^{(1,p-1)} = J ,
\end{eqnarray*}
for $\ell = 1,2, \ldots , (L-1)$, where $t \geq 0$, $J \geq 0$ and $\gamma > 0$ are real.
All other matrix elements are zero, and we assume open boundary conditions.

As examples, the Hermitian SSH model~\cite{ssh79,ssh80,asboth16}, Fig.~\ref{nhfig1}(a), corresponds to $p=2$,
\begin{eqnarray}
{\cal H}^{(1,1)} = \begin{pmatrix}
0 & t & 0 & 0 & 0  & \hdots & 0 & 0 & 0 \\
t & 0 & J & 0 & 0  & \hdots & 0 & 0 & 0 \\
0 & J & 0 & t & 0  & \hdots & 0 & 0 & 0 \\
0 & 0 & t & 0 & J  & \hdots & 0 & 0 & 0 \\
0 & 0 & 0 & J & 0  & \hdots & 0 & 0 & 0 \\
\vdots & \vdots & \vdots & \vdots & \vdots & \ddots & \vdots & \vdots & \vdots \\
0 & 0 & 0 & 0 & 0 & \hdots & 0 & J & 0 \\
0 & 0 & 0 & 0 & 0 & \hdots & J & 0 & t \\
0 & 0 & 0 & 0 & 0 & \hdots & 0 & t & 0
\end{pmatrix} .
\end{eqnarray}
For $p=3$, ${\cal H}^{(1,2)}$, Fig.~\ref{nhfig1}(b), is given by
\begin{eqnarray}
{\cal H}^{(1,2)} = \begin{pmatrix}
0 & 0 & t & 0 & 0 & 0 & 0 & \hdots & 0 & 0 & 0 & 0 \\
t & 0 & 0 & J & 0 & 0 & 0 & \hdots & 0 & 0 & 0 & 0 \\
0 & \gamma & 0 & 0 & 0 & 0 & 0 & \hdots & 0 & 0 & 0 & 0 \\
0 & 0 & J & 0 & 0 & t & 0 & \hdots & 0 & 0 & 0 & 0 \\
0 & 0 & 0 & t & 0 & 0 & J & \hdots & 0 & 0 & 0 & 0 \\
0 & 0 & 0 & 0 & \gamma & 0 & 0 & \hdots & 0 & 0 & 0 & 0 \\
0 & 0 & 0 & 0 & 0 & J & 0 & \hdots & 0 & 0 & 0 & 0 \\
\vdots & \vdots & \vdots & \vdots & \vdots & \vdots & \vdots & \ddots & \vdots & \vdots & \vdots & \vdots \\
0 & 0 & 0 & 0 & 0 & 0 & 0 & \hdots  & J & 0 & 0 & t \\
0 & 0 & 0 & 0 & 0 & 0 & 0 & \hdots  & 0 & t & 0 & 0 \\
0 & 0 & 0 & 0 & 0 & 0 & 0 & \hdots  & 0 & 0 & \gamma & 0
\end{pmatrix} . \label{hp3}
\end{eqnarray}
The matrices ${\cal H}^{(1,p-1)}$ are the same as the matrices in the construction~\cite{fendley14} of generalized raising and lowering operators for $p$-state parafermions (denoted ${\cal M}_n$ in~\cite{fendley14}), except for re-scaling of the hopping parameters. Hence, they have the same single-particle energy levels as free parafermions.
In particular, for a system with $L$ unit cells, the SSH model ${\cal H}^{(1,1)}$ has $L$ positive eigenvalues, denoted $\epsilon_j^{\textrm{SSH}}$, and ${\cal H}^{(m,n)}$ has $pL$ eigenvalues $\epsilon_j^{(p)}$ given by
\begin{eqnarray}
\left( \epsilon_j^{(p)} \right)^p = \gamma^{p-2} \left( \epsilon_j^{\textrm{SSH}} \right)^2 , \label{clock}
\end{eqnarray}
where $j = 1,2,\ldots , L$ and $p =m+n$.
Note that, although the non-Hermitian SSH model has the same single-particle energy levels as Baxter's clock model~\cite{baxter89a,baxter89b,fendley14}, its many-body energy spectrum differs due to different occupation numbers of fermions and parafermions. The models here are constructed using fermionic creation and annihilation operators, and they do not exhibit the unusual exchange and braiding properties of parafermions~\cite{fendley14}.
Nevertheless, as the two models share the same single-particle levels, they have common symmetries and topology, determined by the Hamiltonian matrix~(\ref{hp3}), say, as we describe in the following sections.
Note that the topological properties of the Hamiltonian matrix may be simulated by classical systems such as topolectrical circuits~\cite{hofmann19,helbig20,liu21,rafiulislam21,zhang21b,zou21,xie25,sahin25} 
which generally take no account of occupation numbers and, thus, would not distinguish between the non-Hermitian SSH model and Baxter's clock model.

\section{Partial unidirectional hopping}\label{s:partial}

\subsection{Symmetries}

We now consider systems with partial unidirectional hopping by considering the Hamiltonian
\begin{eqnarray}
\!\!\!\!\!\! \!\!\!\!
{\cal H}^{(m,n)} (k,u) \!=\! {\cal H}^{(m,n)} (k,1) + (1-u) \big( {\cal H}^{(m,n)} (k,1) \big)^{\dagger} \!\! , \label{ugen}
\end{eqnarray}
where ${\cal H}^{(m,n)} (k,1)$ is the fully unidirectional Bloch Hamiltonian  defined in Eq.~(\ref{hgen}) and the degree of directionality $0 \leq u \leq 1$. Hamiltonian ${\cal H}^{(m,n)} (k,u)$ is Hermitian for $u=0$ and non-Hermitian otherwise.

We begin by describing the symmetries of ${\cal H}^{(m,n)} (k,u)$ using the definitions of the symmetry classification for non-Hermitian Hamiltonians~\cite{kawabata19a,kawabata19b}. Throughout this paper, we consider only real tight-binding parameters, so ${\cal H}^{(m,n)} (k,u)$ satisfies time-reversal symmetry,
\begin{eqnarray}
\big( {\cal H}^{(m,n)} (k,u) \big)^{\ast} = {\cal H}^{(m,n)} (-k,u) , \label{trs}
\end{eqnarray}
for all $u$ values.
This dictates that the energy spectrum is either real or comes with complex-conjugate pairs~\cite{kawabata19a} (it has reflection symmetry in the real energy axis).
For fully unidirectional hopping, $u=1$, the Hamiltonian also satisfies complex chiral symmetry~(\ref{chiralk}) and this guarantees that the spectrum has the form~(\ref{pdis2}) given by the $p$th roots of unity and the parent model.
The combination of time-reversal symmetry and chiral symmetry~(\ref{chiralk}) gives a generalized particle-hole symmetry $Z \big( {\cal H}^{(m,n)} (k,1) \big)^{\ast} Z^{-1} = \omega {\cal H}^{(m,n)} (-k,1)$.

For partial unidirectional hopping, $0 < u < 1$, the symmetries depend on whether the number of orbitals per cell $p = m+n$ is even or odd.
For odd $p$, the Hamiltonian is pseudo-Hermitian $\eta \big( {\cal H}^{(m,n)} (k,u) \big)^{\dagger} \eta^{-1} = {\cal H}^{(m,n)} (k,u)$ where $\eta$ is unitary and Hermitian~\cite{mostafazadeh02,kawabata19b}.
Combined with time-reversal symmetry, this gives a variant time-reversal symmetry, ${\cal C}_{+} \big( {\cal H}^{(m,n)} (k,u) \big)^{T} {\cal C}_{+}^{-1} = {\cal H}^{(m,n)} (-k,u)$ where ${\cal C}_{+} {\cal C}_{+}^{\ast} = +1$~\cite{kawabata19b}.
For even $p$, the Hamiltonian obeys sublattice symmetry
$S {\cal H}^{(m,n)} (k,u) S^{-1} = - {\cal H}^{(m,n)} (k,u)$ where $S^2 = 1$~\cite{kawabata19b}.
Combined with time-reversal symmetry, this gives a variant particle-hole symmetry,
${\cal T}_{-} \big( {\cal H}^{(m,n)} (k,u) \big)^{\ast} {\cal T}_{-}^{-1} = - {\cal H}^{(m,n)} (-k,u)$ where ${\cal T}_{-} {\cal T}_{-}^{\ast} = +1$~\cite{kawabata19b}.
This dictates that the energy spectrum for even $p$ is either purely imaginary or comes with $( \epsilon , -\epsilon^{\ast} )$ pairs~\cite{kawabata19a} (it has reflection symmetry in the imaginary energy axis).
The topology of the models, resulting from the symmetries, is discussed in Sec.~\ref{s:top}.

For bidirectional hopping, ${\cal H}^{(m,n)} (k,0)$ is an Hermitian Hamiltonian.
It can be block diagonalized into even and odd parity blocks with respect to inversion of the orbitals within the unit cell.
Partially unidirectional hopping $u > 0$ breaks the inversion symmetry and mixes the even and odd blocks, and the real energy spectrum evolves into a complex one as the degree of unidirectionality $u$ increases.
This process is determined by the topology of the parent model and by the number of orbitals per unit cell, $p = m + n$.
Below, we describe this process in detail for $p=3$ and $p=4$ with the example of the SSH model.

\begin{figure*}[t]
\includegraphics[scale=0.48]{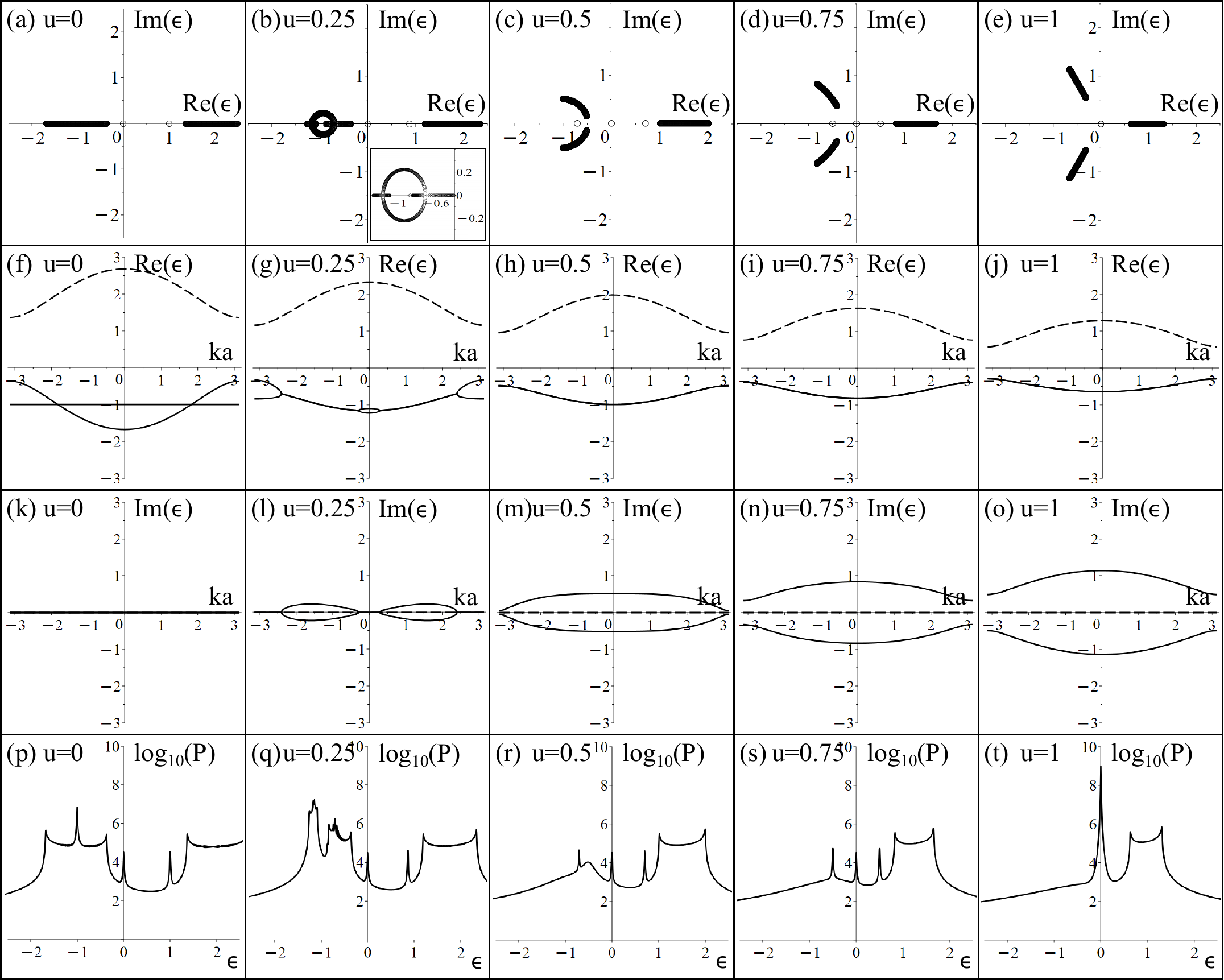}
\caption{Complex energy spectra for $p=3$ orbitals, model $H^{(1,2)}$, as a function of the degree of unidirectionality $u$ when the SSH model is the parent model.
The top row shows energy eigenvalues (circles) determined numerically in position space by diagonalizing ${\cal H}^{(1,2)} (u) = {\cal H}^{(1,2)} (1) + (1-u) \big( {\cal H}^{(1,2)} (1) \big)^{\dagger}$ where ${\cal H}^{(1,2)} (1)$ is given in Eq.~(\ref{hp3}), using open boundary conditions and $L = 200$ unit cells.
The inset in (b) shows a close up of the region where the eigenvalues form a circular shape. For all $u$ values, there are three edge states with energies on the real axis (isolated circles, except one is obscured by other energies in (a) and (b)), and they are threefold degenerate at zero energy for $u=1$ in (e).
The second row shows the real part of the energy bands and the third row shows their imaginary part, plotted for $-\pi \leq k a \leq \pi$ and obtained by diagonalizing the Bloch Hamiltonian~(\ref{blochh12}).
Dashed lines show the band which is always real, and solid lines show the two bands which are partly real and partly imaginary.
When the imaginary parts of the latter (solid lines) are non-zero, their real parts are superimposed on each other and appear as a single line in the plots.
The bottom row shows the response power $P(\epsilon)$~(\ref{powerdef}) as a function of real energy $\epsilon$ determined in position space using open boundary conditions and $L = 200$ unit cells.
To smooth these plots, we add a small imaginary energy as $\epsilon + i \delta$ where $\delta = 0.005$.
In all plots, parameter values are $t = 0.5$ and $J = \gamma = 1.0$.
}\label{p3fig}
\end{figure*}

\subsection{$H^{(1,2)}$}\label{s:h12}

There is one model with $p=3$, $H^{(1,2)}$, with one orbital on the $A$ site and two on the $B$ site, Fig.~\ref{nhfig1}(b). The Bloch Hamiltonian ${\cal H}^{(1,2)} (k,1)$ written in the $A$, $B1$, $B2$ basis is given by Eq.~(\ref{h12u1}).
In the bidirectional limit, there is a symmetry related to swapping the $B1$ and $B2$ orbitals. Hence, we write the Hamiltonian ${\cal H}^{(1,2)} (k,u)$ in a basis of even and odd parity states, $A$, $(B1 + B2)/\sqrt{2}$, $(B1 - B2)/\sqrt{2}$, as
\begin{eqnarray}
\!\!\!\!\!
{\tilde{\cal H}}^{(1,2)} (k,u) = \begin{pmatrix}
0 & \sqrt{2} \tilde{h}^{\ast} (k) & - \sqrt{2} \tilde{u}\tilde{h}^{\ast} (k) \\
\sqrt{2} \tilde{h} (k) & \tilde{\gamma} & \tilde{u} \tilde{\gamma} \\
\sqrt{2} \tilde{u}\tilde{h} (k) & - \tilde{u} \tilde{\gamma} & - \tilde{\gamma}
\end{pmatrix} \!\! , \label{h12u}
\end{eqnarray}
where
\begin{eqnarray}
\tilde{h} (k) &=& (1 - u/2) h (k) , \label{htilde} \\
\tilde{\gamma} &=& (1 - u/2) \gamma , \label{gtilde} \\
\tilde{u} &=& u / (2-u) . \label{utilde}
\end{eqnarray}
This explicitly illustrates that the unidirectional hopping $u$ breaks the inversion symmetry and mixes the even and odd blocks.

The odd state gives a flat band~\cite{leykam13,flach14,leykam18,kremer20,marques21,danieli24}, and the even $2 \times 2$ block takes a form similar to the parent model, but with different onsite energies (this is the Rice-Mele model~\cite{ricemele82} if the parent Hamiltonian is the SSH model).
The energy eigenvalues of the even and odd blocks on their own are
\begin{eqnarray}
E_{1,2} (k) &=& \frac{\tilde{\gamma}}{2} \pm \sqrt{ \frac{\tilde{\gamma}^2}{4} + 2 | \tilde{h} |^2 } , \label{evenh12} \\
E_3 &=& - \tilde{\gamma} . \label{oddh12}
\end{eqnarray}
Now we write the Hamiltonian in the eigenbasis of these states,
\begin{eqnarray}
{\bar{\cal H}}^{(1,2)} (k,u) = \begin{pmatrix}
E_1 & 0 & - \tilde{u} a_1 \\
0 & E_2 & - \tilde{u} a_2 \\
 \tilde{u} a_1 & \tilde{u} a_2 & E_3
\end{pmatrix} ,  \label{blochh12}
\end{eqnarray}
where
\begin{eqnarray}
a_{1,2} = \frac{2 \left( | \tilde{h} |^2 - \tilde{\gamma} E_{1,2} / 2  \right)}{\sqrt{E_{1,2}^2 + 2| \tilde{h} |^2}} .
\end{eqnarray}
Depending on the particular form of $h(k)$, energies $E_2$ and $E_3$ are degenerate or nearly degenerate for some $k$ values.
For small $u$, we describe the mixing of these degenerate states using a $2 \times 2$ effective Hamiltonian,
\begin{eqnarray}
{\cal H}_{\mathrm{eff}}^{(1,2)} (k,u) = \begin{pmatrix}
E_2 & - \tilde{u} a_2 \\
 \tilde{u} a_2 & - \tilde{\gamma}
\end{pmatrix} ,
\end{eqnarray}
which has energies
\begin{eqnarray}
\epsilon_{\pm} (k) = \frac{E_2 - \tilde{\gamma}}{2} \pm \sqrt{ \frac{\left( E_2 + \tilde{\gamma} \right)^2}{4} -  \tilde{u}^2 a_2^2} . \label{circles}
\end{eqnarray}
As $u$ increases, the band energies $\epsilon_{\pm} (k)$ evolve from being purely real to being complex.
Since $|h(k)|$ varies across the band, the evolution is done via the formation of a circular band structure, centered on the real axis.

For the SSH model, $h(k) = t + J e^{ika}$. Across the first Brillouin zone, the maximum value of $|h(k)|$ is $|t+J|$ and the minimum value is $|t-J|$.
The evolution of the complex energy spectra as a function of the degree of unidirectionality $u$ is shown in Fig.~\ref{p3fig} (top row). Energy eigenvalues are determined numerically in position space with open boundary conditions by diagonalizing ${\cal H}^{(1,2)} (u) = {\cal H}^{(1,2)} (1) + (1-u) \big( {\cal H}^{(1,2)} (1) \big)^{\dagger}$ where ${\cal H}^{(1,2)} (1)$ is given in Eq.~(\ref{hp3}). Parameter values are $t = 0.5$, $J = \gamma = 1.0$, and there are $L = 200$ unit cells.
In Fig.~\ref{p3fig}, the second row shows the real part of the energy bands and the third row shows their imaginary part, plotted for $-\pi \leq k a \leq \pi$ and obtained by diagonalizing the Bloch Hamiltonian~(\ref{blochh12}).

For $u=0$, Fig.~\ref{p3fig}(a), the even parity blocks given two real bands equivalent to the Rice-Mele model~\cite{ricemele82} with a band gap centered on energy $\gamma / 2$~(\ref{evenh12}).
In addition, the odd parity states give a flat band at energy $-\gamma$~(\ref{oddh12}).
For $u=0.25$, Fig.~\ref{p3fig}(b), the two bands with $\mathrm{Re} (\epsilon) < 0$ have some purely real energies and some complex ones forming a circular structure according to Eq.~(\ref{circles}).
For $u=0.5$, Fig.~\ref{p3fig}(c), these two bands contain no purely real energy values, and they continue to evolve until $u=1$, Fig.~\ref{p3fig}(e), where they are described by Eq.~(\ref{pdis3}).
The details of how the two bands with $\mathrm{Re} (\epsilon) < 0$ coalesce, according to Eq.~(\ref{circles}), depend on the bandwidth of $|h(k)|$. Plots with different values of $t$ and $J$ are given in the Supplemental Material~\cite{supplementary}.

With $t < J$, the parent SSH model is in the topologically nontrivial phase.
For all $u$ values, there are edge states with real energies, shown in Fig.~\ref{p3fig} (top row). Their energies may be estimated by considering the trimer limit $t=0$, Fig.~\ref{nhfig1}(d), in which there are three edge states: one on the left side has energy $\epsilon = 0$ and two on the right side have energies $\epsilon = \pm \gamma \sqrt{1 - u}$. These estimates are a good approximation for $J > t > 0$ beyond the trimer limit in a large enough system, and they are in excellent agreement with the numerical data in Fig.~\ref{p3fig}.

In general, the edge states give defective eigenvalues when $u=1$, $t = 0$. They are at zero energy with an algebraic multiplicity of $p$ and a geometric multiplicity of two with partial degeneracies~\cite{bid25a,bid25b} $(l_1 , l_2 ) = (m,n)$, partitioned according to the left and right ends of the system.
The defective eigenvalues produce a characteristic resonant response determined by the largest partial degeneracy $\ell_{\mathrm{m}} = \mathrm{max} (m,n)$ of the fragmented exceptional point~\cite{bid25a,bid25b}. 
With the Green's function $G(\epsilon) = ( \epsilon \mathbb{I} - {\cal H}^{(m,n)} )^{-1}$, the spectrally-resolved response power~\cite{bid25a,kullig25,bid25b} is
\begin{eqnarray}
P(\epsilon) = \mathrm{tr} \{ [  G(\epsilon) ]^{\dagger} G(\epsilon) \} . \label{powerdef}
\end{eqnarray}
For the edge states, we find that
\begin{eqnarray}
P(\epsilon) \sim \frac{\gamma^{2\ell_{\mathrm{m}}-2}}{|\epsilon|^{2\ell_{\mathrm{m}}}} ;
\qquad \ell_{\mathrm{m}} = \mathrm{max} (m,n) , \label{response}
\end{eqnarray}
for $u=1$, $t = 0$, and energy $\epsilon$ near zero, in agreement with Refs.~\cite{bid25a,bid25b}.
The power~(\ref{powerdef}) as a function of the degree of unidirectionality $u$ is plotted for $H^{(1,2)}$ in Fig.~\ref{p3fig} (bottom row) for $J > t > 0$, determined numerically in position space with open boundary conditions. There are plateaus at the location of the bands (on the real energy axis) and distinctive peaks corresponding to the edge states, with the most prominent peak appearing at zero energy for $u=1$. Similar peaks will appear due to states localized on domain walls, as described in Sec.~\ref{s:soliton}.

\begin{figure}[t]
\includegraphics[scale=0.50]{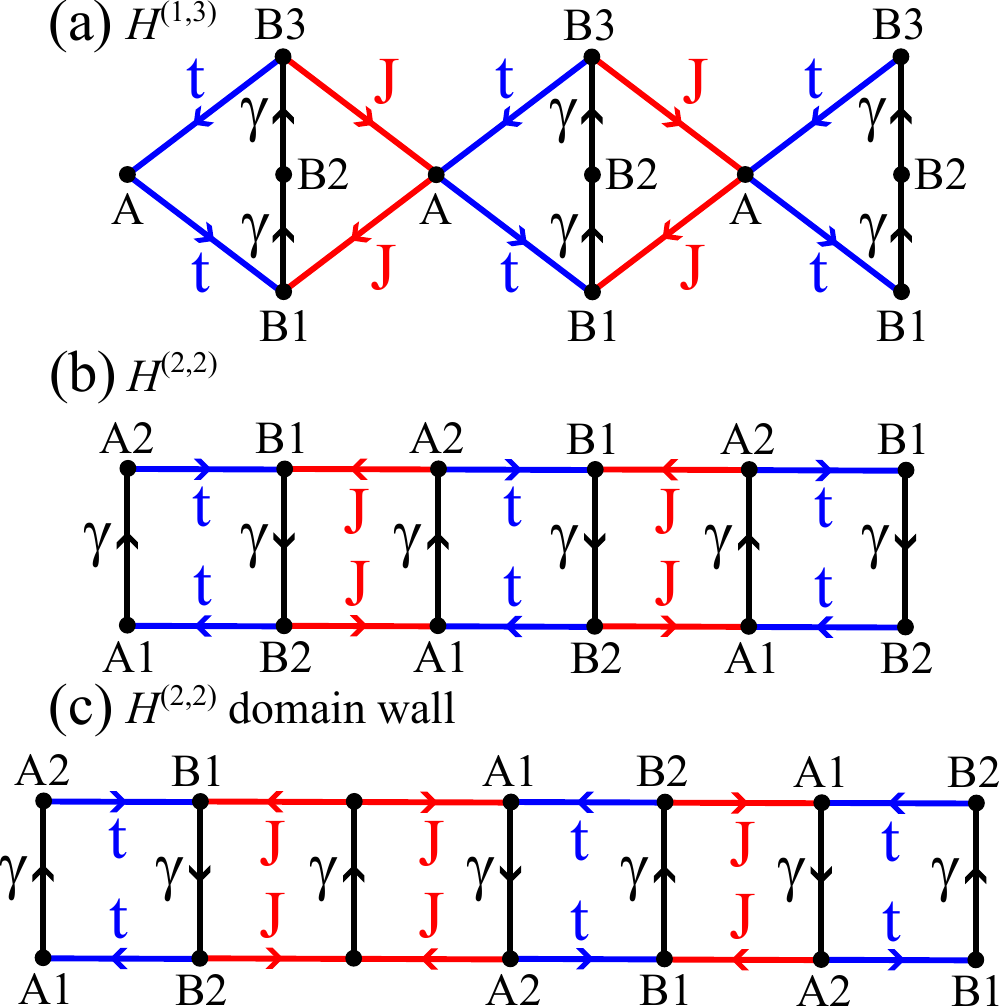}
\caption{Non-Hermitian Hamiltonians formed from the parent SSH model with $p=4$ orbitals per unit cell.
(a) Model $H^{(1,3)}$ with three orbitals $B1$, $B2$, $B3$ connected by unidirectional hopping $\gamma > 0$ as indicated by the arrows, with unidirectional hopping from $A$ to $B1$ and from $B3$ to $A$.
(b) Model $H^{(2,2)}$ with two orbitals $A1$, $A2$ and two orbitals $B1$, $B2$ connected by unidirectional hopping $\gamma > 0$, with unidirectional hopping from $A2$ to $B1$ and from $B2$ to $A1$.
(c) Model $H^{(2,2)}$ with a domain wall hosting localized states.
}\label{p4models}
\end{figure}

\begin{figure*}[t]
\includegraphics[scale=0.48]{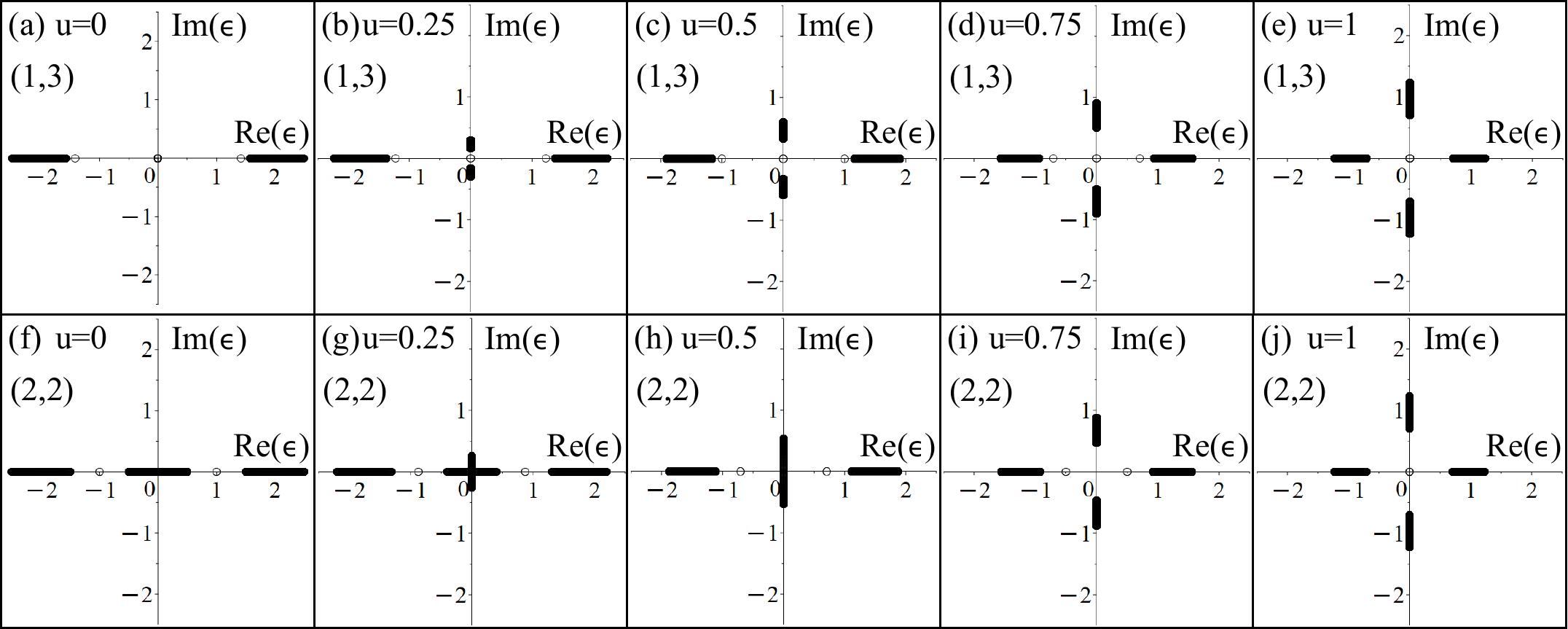}
\caption{Complex energy spectra for $p=4$ orbitals, as a function of the degree of unidirectionality $u$ when the SSH model is the parent model.
The top row shows model $H^{(1,3)}$, the bottom row $H^{(2,2)}$.
Energy eigenvalues (circles) are determined numerically in position space with open boundary conditions by diagonalizing the Hamiltonian. Parameter values are $t = 0.5$, $J = \gamma = 1.0$, and there are $L = 200$ unit cells.
For all $u$ values, there are four edge states with energies on the real axis (isolated circles). For $H^{(1,3)}$ (top row), two are degenerate at zero energy, the other two are at non-zero energy for $u < 1$. 
For $H^{(2,2)}$ (bottom row), the edge states are twofold degenerate for all $u < 1$. For $u=1$, the edge states are fourfold degenerate at zero energy in (e) and (j).}\label{p4fig}
\end{figure*}

\subsection{$H^{(1,3)}$}

There are two models with $p=4$, $H^{(1,3)}$ and $H^{(2,2)}$: $H^{(1,3)}$ has one orbital on the $A$ site and three on the $B$ site, Fig.~\ref{p4models}(a).
In the bidirectional limit, there is a symmetry related to swapping the $B1$ and $B3$ orbitals. Hence, we write the Hamiltonian ${\cal H}^{(1,3)} (k,u)$ in a basis of even and odd parity states, $A$, $(B1 + B3)/\sqrt{2}$, $B2$, $(B1 - B3)/\sqrt{2}$, as
\begin{eqnarray}
\!\!\!\!\! {\tilde{\cal H}}^{(1,3)} (k,u) = \sqrt{2} \begin{pmatrix}
0 & \tilde{h}^{\ast} (k) & 0 & - \tilde{u} \tilde{h}^{\ast} (k) \\
\tilde{h} (k) & 0 & \tilde{\gamma} & 0 \\
0 & \tilde{\gamma} & 0 & \tilde{u} \tilde{\gamma} \\
\tilde{u} \tilde{h} (k) & 0 & - \tilde{u} \tilde{\gamma} & 0
\end{pmatrix} \!\! , \label{h13u}
\end{eqnarray}
using the definitions in Eqs.~(\ref{htilde})-(\ref{utilde}).
The even $3 \times 3$ block has two dispersive bands plus a flat band at zero energy. The odd state also gives a flat band at zero energy.
The energy eigenvalues of the even and odd blocks on their own are
\begin{eqnarray}
E_{1,2} (k) &=& \pm \sqrt{ 2 \left( \tilde{\gamma}^2 + | \tilde{h} |^2 \right)} , \label{h13e12} \\
E_3 = E_4 &=& 0 , \label{h13e34}
\end{eqnarray}
and the Hamiltonian in the eigenbasis of these states is
\begin{eqnarray}
{\bar{\cal H}}^{(1,3)} (k,u) = \begin{pmatrix}
E_1 & 0 & 0 & \tilde{u} b_1 \\
0 & E_2 & 0 & \tilde{u} b_1 \\
0 & 0 & 0 & \tilde{u} b_2^{\ast} \\
- \tilde{u} b_1 & - \tilde{u} b_1 & - \tilde{u} b_2 & 0
\end{pmatrix} ,
\end{eqnarray}
where
\begin{eqnarray}
b_1 = \frac{\left( \tilde{\gamma}^2 - | \tilde{h} |^2 \right) }{\sqrt{\tilde{\gamma}^2 + | \tilde{h} |^2}} , \qquad
b_2 = \frac{2 \sqrt{2} \tilde{\gamma} \tilde{h}}{\sqrt{\tilde{\gamma}^2 + | \tilde{h} |^2}} .
\end{eqnarray}
For small $u$, we consider a $2 \times 2$ effective Hamiltonian describing mixing of the two zero-energy flat bands,
\begin{eqnarray}
{\cal H}_{\mathrm{eff}}^{(1,3)} (k,u) = \begin{pmatrix}
0 & \tilde{u} b_2^{\ast} \\
- \tilde{u} b_2 & 0
\end{pmatrix} ,
\end{eqnarray}
which has energies
\begin{eqnarray}
\epsilon_{\pm} (k) = \pm \frac{2 \sqrt{2} i \tilde{u} \tilde{\gamma} |\tilde{h} (k)|}{\sqrt{ \tilde{\gamma}^2 + | \tilde{h} (k) |^2}} . \label{h13imag}
\end{eqnarray}
Hence these two bands are purely imaginary for $u > 0$ (and $| \tilde{h} (k) | \neq 0$).

The evolution of the complex energy spectra as a function of the degree of unidirectionality $u$ is shown in Fig.~\ref{p4fig} (top row). Energy eigenvalues are determined numerically in position space with open boundary conditions. Parameter values are $t = 0.5$, $J = \gamma = 1.0$, and there are $L = 200$ unit cells.
For all $u$, all energies lie on either the real or the imaginary axes.
For $u=0$, Fig.~\ref{p4fig}(a), energies are given by Eqs.~(\ref{h13e12}) and~(\ref{h13e34}) with two degenerate flat bands at zero energy.
For $u > 0$, Fig.~\ref{p4fig}(b), these two bands become purely imaginary and dispersive, in accordance with Eq.~(\ref{h13imag}).
They remain on the imaginary axis for all subsequent $u$, and, at $u=1$, Fig.~\ref{p4fig}(e), where they are described by Eq.~(\ref{pdis3}).

With $t < J$, there are four edge states with real energies for all $u$ values, shown in Fig.~\ref{p4fig} (top row).
Their energies may be estimated by considering the limit $t=0$ in Fig.~\ref{p4models}(a).
One edge state on the left side has energy $\epsilon = 0$ and three on the right side have energies $\epsilon = 0$ and $\epsilon = \pm \gamma \sqrt{2(1 - u)}$. These estimates are a good approximation for $J > t > 0$ beyond the $t=0$ limit in a large enough system, and they are in excellent agreement with the numerical data in Fig.~\ref{p4fig} (top row).

\begin{figure*}[t]
\includegraphics[scale=0.48]{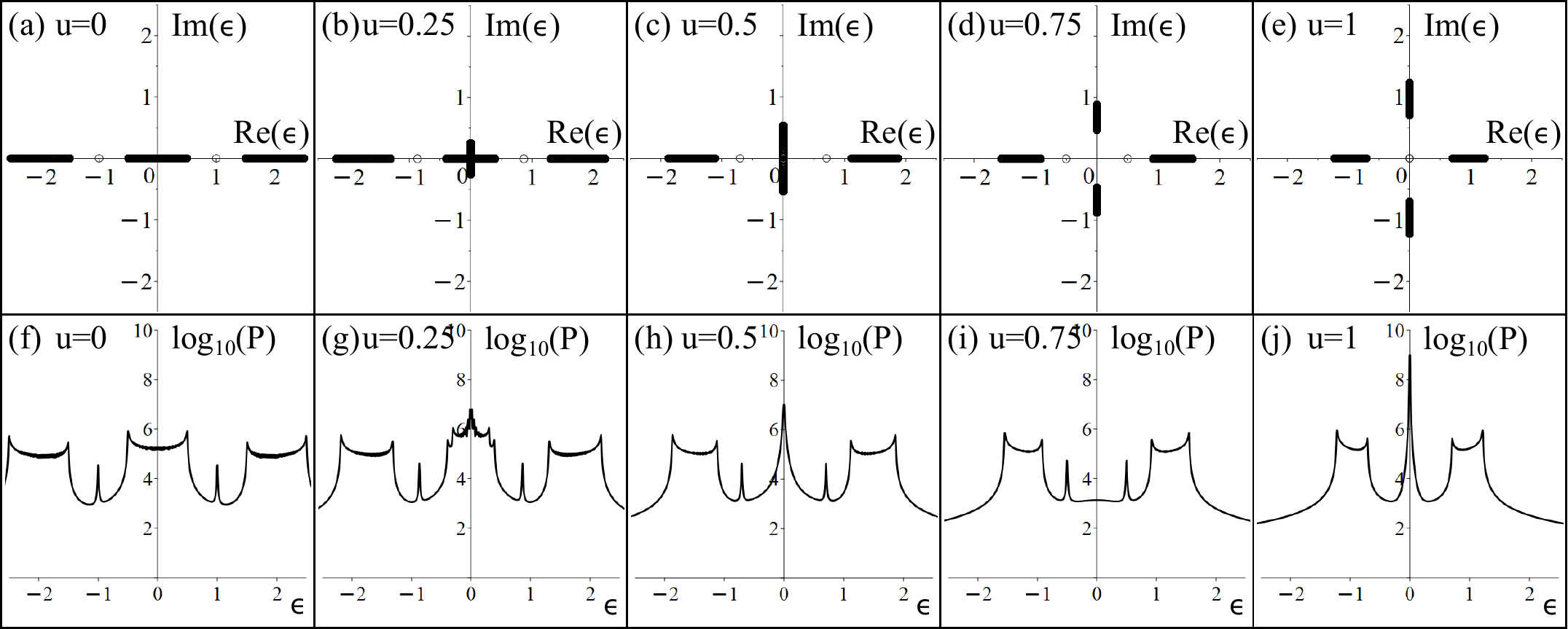}
\caption{Complex energy spectra (top row) and the response power $P(\epsilon)$ (bottom row) with a domain wall in $H^{(2,2)}$ when the SSH model is the parent model.
All data is determined numerically in position space with open boundary conditions by diagonalizing the Hamiltonian. Parameter values are $t = \gamma = 1.0$, $J = 0.5$, and there are $802$ orbitals.
We consider strong $t$ bonds at the edges and a domain wall at the center of the system.
For the energy spectra (top row), states localized on the domain wall are isolated circles on the real energy axis, and they are twofold degenerate at zero energy for $u=1$.
The bottom row shows the response power $P(\epsilon)$~(\ref{powerdef}) as a function of real energy $\epsilon$ where we add a small imaginary energy as $\epsilon + i \delta$ with $\delta = 0.005$.
}\label{solplot1}
\end{figure*}

\subsection{$H^{(2,2)}$}\label{s:h22}

The non-Hermitian model $H^{(2,2)}$ has two orbitals on the $A$ site and two on the $B$ site, Fig.~\ref{p4models}(b).
Unlike $H^{(1,2)}$ and $H^{(1,3)}$, there is parity related to inverting the whole system (e.g. swapping $A1$ on the left edge with $B1$ on the right edge in Fig.~\ref{p4models}(b)); the parent (SSH) model has an analogous symmetry.
In the bidirectional limit, $H^{(2,2)}$ corresponds to two coupled SSH chains~\cite{clay05,li13,cheon15,zhang15,zhang17,li17,matveeva23}, and, as with the other models, there is symmetry related to swapping orbitals within the unit cell ($A1$ and $A2$, and $B1$ and $B2$).
We write the Hamiltonian ${\cal H}^{(2,2)} (k,u)$ in a basis of even and odd parity states, $(A1 + A2)/\sqrt{2}$, $(B1 + B2)/\sqrt{2}$, $(A1 - A2)/\sqrt{2}$, $(B1 - B2)/\sqrt{2}$, as
\begin{eqnarray}
\!\!\!\!\!\! {\tilde{\cal H}}^{(2,2)} (k,u) = \begin{pmatrix}
\tilde{\gamma} & \tilde{h}^{\ast} (k) & \tilde{u} \tilde{\gamma} & - \tilde{u} \tilde{h}^{\ast} (k) \\
\tilde{h} (k) & \tilde{\gamma} & - \tilde{u} \tilde{h} (k) & \tilde{u} \tilde{\gamma} \\
-\tilde{u} \tilde{\gamma} & \tilde{u} \tilde{h}^{\ast} (k) & - \tilde{\gamma} & -\tilde{h}^{\ast} (k) \\
\tilde{u} \tilde{h} (k) & -\tilde{u} \tilde{\gamma} & -\tilde{h} (k) & - \tilde{\gamma}
\end{pmatrix} \!\! , \label{h22u}
\end{eqnarray}
using the definitions in Eqs.~(\ref{htilde})-(\ref{utilde}).
The even $2 \times 2$ block takes the form of the parent model centered on energy $\tilde{\gamma}$ and the odd $2 \times 2$ block takes the form of the parent model centered on energy $- \tilde{\gamma}$.
The energy eigenvalues of the even and odd blocks on their own are
\begin{eqnarray}
E_{1} (k) &=& \tilde{\gamma} + | \tilde{h} | ; \qquad E_{2} (k) = \tilde{\gamma} - | \tilde{h} | \label{h22e12} \\
E_{3} (k) &=& - \tilde{\gamma}+ | \tilde{h} | ; \qquad E_{4} (k) = - \tilde{\gamma} - | \tilde{h} | , \label{h22e34}
\end{eqnarray}
and the Hamiltonian in the eigenbasis of these states may be written in a block diagonal form as
\begin{eqnarray}
{\bar{\cal H}}^{(2,2)} (k,u) = \begin{pmatrix}
E_1 & \tilde{u} E_2 & 0 & 0 \\
- \tilde{u} E_2 & E_4 & 0 & 0 \\
0 & 0 & E_2 & \tilde{u} E_1 \\
0 & 0 & -\tilde{u} E_1 & E_3
\end{pmatrix} .
\end{eqnarray}
with energies
\begin{eqnarray}
\epsilon_{\mathrm{r},\pm} (k) &=& \pm \sqrt{ \big( \tilde{\gamma} + | \tilde{h} | \big)^2 - \tilde{u}^2 \big( \tilde{\gamma} - | \tilde{h} | \big)^2} , \label{h22rt} \\
\epsilon_{\mathrm{i},\pm} (k) &=& \pm \sqrt{ \big( \tilde{\gamma} - | \tilde{h} | \big)^2 - \tilde{u}^2 \big( \tilde{\gamma} + | \tilde{h} | \big)^2} . \label{h22it}
\end{eqnarray}
In terms of the original parameters, Eqs.~(\ref{htilde})-(\ref{utilde}),
\begin{eqnarray}
\epsilon_{\mathrm{r},\pm} (k) &=& \pm \sqrt{ \big( \gamma + [1-u] |h| \big) \big( [1-u] \gamma + |h| \big) } , \label{h22r} \\
\epsilon_{\mathrm{i},\pm} (k) &=& \pm \sqrt{ \big( \gamma - [1-u] |h| \big) \big( [1-u] \gamma - |h| \big) } . \label{h22i}
\end{eqnarray}
Thus, there are two bands with real energies $\epsilon_{\mathrm{r},\pm} (k)$ and two bands $\epsilon_{\mathrm{i},\pm} (k)$ with energies that evolve from being purely real to being purely imaginary as $u$ increases.
For $|h(k)| < \gamma$, these bands are imaginary for $|h(k)| > (1-u) \gamma$.

The evolution of the complex energy spectra as a function of the degree of unidirectionality $u$ is shown in Fig.~\ref{p4fig} (bottom row). Energy eigenvalues are determined numerically in position space with open boundary conditions. Parameter values are $t = 0.5$, $J = \gamma = 1.0$, and there are $L = 200$ unit cells.
For all $u$, all energies lie on either the real or the imaginary axes.
For $u=0$, Fig.~\ref{p4fig}(f), the energies are those of two parent (SSH) models centered on $\gamma$ and $-\gamma$ as in Eqs.~(\ref{h22e12}) and~(\ref{h22e34}).
For $u > 0$, Fig.~\ref{p4fig}(g), two bands become partly real and partly imaginary, Eq.~(\ref{h22i}). For large enough $u$, Fig.~\ref{p4fig}(h), they become wholly imaginary and, at $u=1$, Fig.~\ref{p4fig}(j), they are described by Eq.~(\ref{pdis3}).

With $t < J$, there are four edge states with real energies for all $u$ values, shown in Fig.~\ref{p4fig} (bottom row).
Their energies may be estimated by considering the limit $t=0$ in Fig.~\ref{p4models}(b).
Two edge states on the left side have energies $\epsilon = \pm \gamma \sqrt{(1 - u)}$ and they are degenerate with two on the right side, $\epsilon = \pm \gamma \sqrt{(1 - u)}$.
These estimates are a good approximation for $J > t > 0$ beyond the $t=0$ limit in a large enough system, and they are in excellent agreement with the numerical data in Fig.~\ref{p4fig} (bottom row).

The Bloch Hamiltonian ${\cal H}^{(n,n)} (k,1)$ is an $n$th root of the parent Hamiltonian ${\cal H}^{(1,1)} (k,0)$.
With Eq.~(\ref{h22u}), $\big( {\tilde{\cal H}}^{(2,2)} (k,u) \big)^2$ is block diagonal, consisting of two $2 \times 2$ blocks, each of which is
\begin{eqnarray}
\!\!\!\!\!\!\!\!
H_{\mathrm{sq}} (u) = \begin{pmatrix}
(1-u)\big( \gamma^2 + |h|^2 \big) & \big( u^2 - 2u + 2) \gamma h^{\ast} \\
\big( u^2 - 2u + 2) \gamma h & (1-u)\big( \gamma^2 + |h|^2 \big)
\end{pmatrix} \!\! .
\end{eqnarray}
This has the parent model on the off-diagonal with an additional dispersing term on the diagonal, and, for $u=1$, the diagonal term is zero.

\subsection{Topology}\label{s:top}

According to the non-Hermitian symmetry classification of Ref.~\cite{kawabata19}, which describes 38 different symmetry classes, ${\cal H}^{(m,n)}$ for $p = m+n \geq 3$ is in the AI non-Hermitian class owing to the presence of time-reversal symmetry~(TRS).
For even $p$, there is also sublattice symmetry commuting with TRS, giving a $\mathbb{Z}$ topological index in one dimension, but no topology in two dimensions~\cite{kawabata19}. 
For odd $p$, there is also pseudo-Hermiticity commuting with TRS, giving a $\mathbb{Z}$ topological index in one dimension with an imaginary line gap (appropriate for TRS), but no topology in two dimensions~\cite{kawabata19}.

The topology of the non-Hermitian models is determined by the parent model owing to the presence of $k$-independent $\gamma$ factors in the Hamiltonians, as we show explicitly in the Supplemental Material~\cite{supplementary}. In one dimension, assuming the system is gapless, the topology is described by the Hermitian winding number $W$~\cite{asboth16} of the parent model~(\ref{parent}),
\begin{eqnarray}
W = \frac{1}{2\pi i} \int_{-\pi / a}^{\pi / a} dk \frac{d}{dk} \ln h(k) .
\end{eqnarray}
Hence, for the SSH model as parent (which only has two insulating phases due to the limited number of parameters), there is a trivial phase for $t > J$ (winding number of zero), a gapless phase for $t = J$, and a topologically nontrivial phase for $t < J$ (winding number of one).

Note that, for $u \alt 1$, the model ${\cal H}^{(m,n)} (k,u)$ is adiabatically connected to the model for $u=1$, ${\cal H}^{(m,n)} (k,1)$, i.e., the value of $u$ can be increased to $u=1$ without breaking any relevant symmetry or closing a band gap (the classification~\cite{kawabata19} does not include the complex chiral and charge-conjugation symmetries described above). Examples are Fig.~\ref{p4fig}(b) ($u=0.25$) which is connected to Fig.~\ref{p4fig}(e) ($u=1$) and Fig.~\ref{p4fig}(i) ($u=0.75$) which is connected to Fig.~\ref{p4fig}(j) ($u=1$).
However, for $u \alt 1$, the model ${\cal H}^{(m,n)} (k,u)$ is not adiabatically connected to the Hermitian model for $u=0$, ${\cal H}^{(m,n)} (k,0)$, because a band gap must close to reach $u=0$. Examples where the gap has closed are Fig.~\ref{p4fig}(a) ($u=0$) and Fig.~\ref{p4fig}(h) ($u=0.5$). Additional plots illustrating this behavior for other parameter values are given in the Supplemental Material~\cite{supplementary}.

\subsection{Localized states and exceptional points on solitons}\label{s:soliton}

When the SSH model is the parent model, there are edge states with real energies for $t < J$ and, for $u=1$, $t=0$, the edges support defective eigenvalues at zero energy.
Similar behavior occurs for states localized on solitons (domain walls in the relative strength of $t$ and $J$), as we now discuss.
As an example, we consider $H^{(2,2)}$ in position space with open boundary conditions, with a domain wall as shown in Fig.~\ref{p4models}(c).
For clarity, we consider $t > J$ and strong $t$ bonds at the edges (as shown in Fig.~\ref{p4models}(c)) so that the only localized states occur on the soliton, not at the edges~\cite{inui94,allen22}.
The energies of the states localized on the domain wall may be estimated by considering the limit $J=0$. In this case, the energies are $\epsilon = \pm \gamma \sqrt{1-u}$.
For $u=1$, these are defective eigenvalues at zero energy with an algebraic multiplicity of two and a geometric multiplicity of one.
As there are no other zero-energy states (at the ends, say), this is an exceptional point as opposed to the fragmented exceptional point~\cite{bid25a,bid25b} in the system with two edge states, Sec.~\ref{s:h12}.

Numerical results are shown in Fig.~\ref{solplot1} obtained by diagonalizing the Hamiltonian in position space with open boundary conditions, $802$ orbitals in total, and a domain wall at the center of the system. The complex energy spectra, top row of Fig.~\ref{solplot1}, look very similar to those of a fault-free system, bottom row of Fig.~\ref{p4fig}, the main difference is that the isolated energy levels are due to localized states on the domain wall instead of the ends.
The bottom row of Fig.~\ref{solplot1} shows the response power~(\ref{powerdef}) as a function of $u$. There are plateaus at the location of the bands (on the real energy axis) and distinctive peaks corresponding to the edge states, with the most prominent peak appearing at zero energy for $u=1$, similar to the behavior for edge states discussed in Sec.~\ref{s:h12}.

\section{Graphene}\label{s:graphene}

We consider monolayer graphene~\cite{novoselov04,castroneto09,mccann13} as noninteracting fermions on the honeycomb lattice with nearest-neighbor hopping. This is another example of a parent model on a bipartite lattice with chiral symmetry, with the form of Eq.~(\ref{parent}) where we replace $h(k)$ with $h ({\bf k}) = - \gamma_0 f^{\ast}({\bf k})$~\cite{saito98,castroneto09,mccann13} where $\gamma_0$ is the nearest-neighbor hopping parameter, ${\bf k} = (k_x , k_y )$ is a two-dimensional wave vector, and
\begin{eqnarray}
f ({\bf k}) = e^{i k_y a / \sqrt{3}} + 2 e^{-i k_y a / (2\sqrt{3})} \cos (k_x a / 2) , 
\end{eqnarray}
where $a$ is the lattice constant.

We focus on the case of $H^{(2,2)}$, discussed for the SSH model in Sec.~\ref{s:h22}, which has two orbitals ($A1$ and $A2$) on every $A$ site of the honeycomb lattice and two orbitals ($B1$ and $B2$) on every $B$ site. This is similar to Fig.~\ref{p4models}(b) where one can view the top and bottom layers as being honeycomb lattices, with hoppings $t$ and $J$ replaced by $\gamma_0$, and $\gamma$ is an interlayer coupling appearing on every site.
For an arbitrary degree of unidirectionality, $0 \leq u \leq 1$, we may use the formulae in Sec.~\ref{s:h22}, replacing $h(k)$ with $h ({\bf k})$.
A major difference as compared to the SSH model is that the band structure of the parent model is gapless whatever the parameter values, i.e., $h ({\bf k}) = 0$ at the Dirac points in graphene.

\begin{figure*}[t]
\includegraphics[scale=0.48]{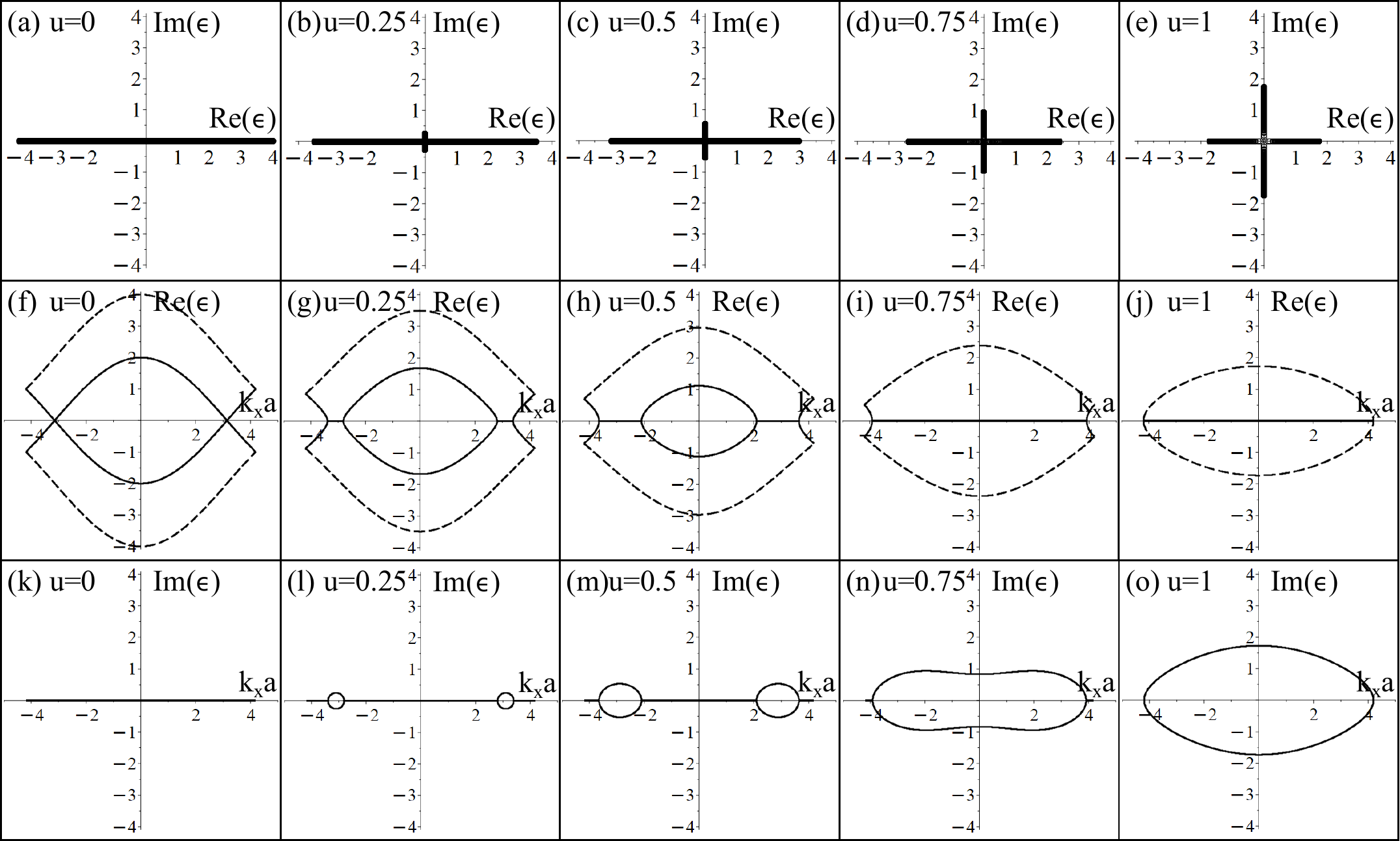}
\caption{Complex energy spectra for $H^{(2,2)}$ with graphene as the parent model, as a function of the degree of unidirectionality $u$.
The top row shows energy eigenvalues determined numerically in position space with periodic boundary conditions and $20000$ orbitals.
The middle row shows the real part of the energy bands and the bottom row shows their imaginary part, plotted for $k_y =0$ and $-4\pi/3 \leq k_x a \leq 4 \pi/3$. Dashed lines show the bands $\epsilon_{\mathrm{r},\pm} (k)$, Eq.~(\ref{h22r}), which are always real, and solid lines show the bands $\epsilon_{\mathrm{i},\pm} (k)$, Eq.~(\ref{h22i}), which are partly real and partly imaginary. For all plots, parameter values are $\gamma = \gamma_0 = 1.0$.
}\label{graphenefig1}
\end{figure*}

The evolution of the complex energy spectra as a function of the degree of unidirectionality $u$ is shown in Fig.~\ref{graphenefig1}. The top row shows energy eigenvalues determined numerically in position space with closed boundary conditions and $20000$ orbitals. The middle row of Fig.~\ref{graphenefig1} shows the real part of the energy bands and the bottom row shows their imaginary part, plotted using analytical formulae: dashed lines show the bands $\epsilon_{\mathrm{r},\pm} ({\bf k})$, Eq.~(\ref{h22r}), which are always real, and solid lines show the bands $\epsilon_{\mathrm{i},\pm} ({\bf k})$, Eq.~(\ref{h22i}), which are partly real and partly imaginary.
The plots are for $k_y =0$ and $-4\pi/3 \leq k_x a \leq 4 \pi/3$. This range includes two Dirac points at 
${\bf K}_{\xi} = \xi (4 \pi/(3a) , 0)$, $\xi = \pm 1$, in the parent monolayer graphene model where $|h ({\bf K}_{\xi}) | = 0$, and, at this point, $\epsilon_{\mathrm{r},\pm} ({\bf K}_{\xi})$ and $\epsilon_{\mathrm{i},\pm} ({\bf K}_{\xi})$ touch.
 For all plots, we use parameter values $\gamma_0 = \gamma = 1.0$ for clarity of the qualitative features in the figures.

For bidirectional hopping $u=0$ (first column in Fig.~\ref{graphenefig1}), $H^{(2,2)}$ is a nearest-neighbor hopping model of $AA$-stacked bilayer graphene~\cite{lee08,rozhkov16} which is Hermitian, and all the energy levels are real as given by Eqs.~(\ref{h22e12}) and (\ref{h22e34}). There are two crossing points at zero energy where $|h ({\bf k}) | = \gamma$.
As $u$ increases, the real bands $\epsilon_{\mathrm{r},\pm} ({\bf k})$ (dashed lines) remain non-zero except for $u=1$ at the points where $|h ({\bf K}_{\xi}) | = 0$.
Bands $\epsilon_{\mathrm{i},\pm} ({\bf k})$ (solid lines) are generally partly real and partly imaginary, except for $u=1$ where they are imaginary (other than the points where $|h ({\bf K}_{\xi}) | = 0$ and the energy is zero).

For $u=1$, the energies are $\epsilon_{\mathrm{r},\pm} ({\bf k}) = \pm \sqrt{\gamma |h ({\bf k}) |}$ and $\epsilon_{\mathrm{i},\pm} ({\bf k}) = \pm i \sqrt{\gamma |h ({\bf k}) |}$.
Monolayer graphene has energies $\pm |h ({\bf k}) |$, and $H^{(2,2)}$ is a square root model~\cite{arkinstall17,pelegri19,kremer20,ezawa20,lin21,marques21,viedma24,li24,huang25,zhao25,song25} of it. Near the Dirac point of graphene, ${\bf k} = {\bf K}_{\xi}$, we consider a small wave vector ${\bf q}$ as ${\bf q} = {\bf k} - {\bf K}_{\xi}$ and $f({\bf k}) \approx - \sqrt{3} a (\xi q_x - i q_y)/2$~\cite{castroneto09,mccann13} giving $h({\bf k}) \approx \hbar v (\xi q_x + i q_y)$ where velocity $v = \sqrt{3} a \gamma_0 / (2\hbar)$. Thus, near the Dirac points, the Bloch Hamiltonian~(\ref{hgen}) is
\begin{eqnarray}
\!\!\!\!\!\!\!\!\!
{\cal H}^{(2,2)} ({\bf q} , 1 ) \approx \begin{pmatrix}
0 & 0 & 0 & Q^{\dagger} \\
\gamma & 0 & 0 & 0 \\
0 & Q & 0 & 0 \\
0 & 0 & \gamma & 0
\end{pmatrix} \!\! ,
\end{eqnarray}
where $Q = \hbar v (\xi q_x + i q_y)$. This has
energies $\epsilon_{\mathrm{r},\pm} ({\bf k}) \approx \pm \sqrt{\gamma \hbar v |{\bf q} |}$ and $\epsilon_{\mathrm{i},\pm} ({\bf k}) \approx \pm i \sqrt{\gamma \hbar v |{\bf q} |}$, in agreement with Eq.~(\ref{pdis2}).
As discussed in Sec.~\ref{s:spectral}, there is an exceptional point~\cite{kato66,heiss04,berry04,klaiman08,heiss12,weimann17,lieu18,kawabata19,miri19,bergholtz21,wiersig22a,wiersig22b,wiersig23,halder23,sayyad23,ryu24,ye24,schomerus24,bid25a,kullig25,bid25b,bid25c} at $|{\bf q} | = 0$ where all eigenvalues are at zero energy with fourfold algebraic multiplicity, but twofold geometric multiplicity.
An experimental signature of the exceptional point is the characteristic resonant response as described by the response power~(\ref{powerdef}), and we find that $P(\epsilon) \sim |\epsilon|^{-4}$ in agreement with Eq.~(\ref{response}) and the expectations of fragmented exceptional points~\cite{bid25a,bid25b}.

As with a trajectory around the Dirac point in the parent model, graphene~\cite{berry84,castroneto09,mccann13}, a trajectory in one of the bands enclosing the exceptional point acquires Berry phase~$\pi$. To show this, we consider right $| \psi_{\mathrm{R},\ell} \rangle$ and left $| \psi_{\mathrm{L},\ell} \rangle$ eigenstates where ${\cal H}^{(2,2)} ({\bf q} , 1 ) | \psi_{\mathrm{R},\ell} \rangle = \epsilon_{\ell} | \psi_{\mathrm{R},\ell} \rangle$, $\big( {\cal H}^{(2,2)} ({\bf q} , 1 ) \big)^{\dagger} | \psi_{\mathrm{L},\ell} \rangle = \epsilon_{\ell}^{\ast} | \psi_{\mathrm{L},\ell} \rangle$, and $\epsilon_{\ell} = i^{\ell} \sqrt{\gamma \hbar v |{\bf q} |}$ for $\ell = 0,1,2,3$. The left and right eigenstates are orthonormal as $\langle \psi_{\mathrm{L},\ell} | \psi_{\mathrm{R},m} \rangle = \delta_{\ell , m}$, and the complex Berry phase $\Upsilon$ may be determined as
$\Upsilon = i \oint \langle \psi_{\mathrm{L},\ell} | \nabla_{\mathbf{q}} |\psi_{\mathrm{R},\ell} \rangle \cdot d \mathbf{q}$~\cite{liang13,shen18,fan20,tsubota22,longhi23}.
Taking a trajectory at constant $|{\bf q} | > 0$ around the exceptional point,
we find $\Upsilon = \pi$ modulo $2\pi$, independent of the band index~$\ell$.

These results can be generalized to $H^{(n,n)}$, the $n$th root model of monolayer graphene with $p=2n$ orbitals.
For $u=1$, and near the Dirac point of graphene, the $p$ energy eigenvalues~(\ref{pdis2}) are given by the complex solutions of
$\big( \epsilon_{\ell}^{(n,n)} (k,1) \big)^{p} = \gamma^{p-2} (\hbar v |{\bf q} |)^2$.
For $|{\bf q}| = 0$, there is an exceptional point as these eigenvalues are at zero energy with $p$-fold algebraic multiplicity, but twofold geometric multiplicity.
The Jordan normal form of the Hamiltonian at the exceptional point is given by the transpose of the Hamiltonian, $\big( {\cal H}^{(n,n)} ({\bf q} = 0 , 1 ) \big)^{T}$, with $\gamma = 1$.
Near the exceptional point, the spectrally-resolved response power~(\ref{powerdef}) behaves as $P(\epsilon) \sim |\epsilon|^{-2n}$.
A trajectory at constant $|{\bf q} | > 0$ around the exceptional point acquires Berry phase $\Upsilon = \pi$ modulo $2\pi$, independent of the band index~$\ell$.

In topolectrical circuits, non-reciprocal hopping is now routinely simulated by non-reciprocal capacitances created using negative impedance converters~\cite{hofmann19,helbig20,liu21,rafiulislam21,zhang21b,zou21,xie25,sahin25},
and the presence of exceptional points has been observed through features in impedance measurements~\cite{helbig20}.
Topolectrical circuits representing graphene and related materials have been proposed~\cite{lee18,luo18,yao22} and realized experimentally~\cite{luo25,xie25}, including a non-Hermitian version of graphene with non-reciprocal hopping~\cite{xie25}.

\section{Generalizations of the construction}

For simplicity, we considered translational invariance of the tight-binding parameters which enables the use of the $k$ space representation, Sec.~\ref{s:general}.
As we consider the parent models to be Hermitian, we also assumed that hopping directed towards the right, e.g., from $A$ to $B1$ in Fig.~\ref{nhfig1}(b), is the same strength as hopping directed towards the left within the same unit cell, e.g., from $B2$ to $A$ in Fig.~\ref{nhfig1}(b). This latter condition ensures that, although the hopping along a single bond is non-reciprocal, hopping is reciprocal when considering the unit cell as a whole. As a result, this construction does not display the non-Hermitian skin effect~\cite{lee16,yao18,song19,liu19b,han21,bergholtz21}.

These conditions may be relaxed, and the model will still satisfy the generalized chiral symmetry~(\ref{chiralk}). For example, the matrix for the SSH model~(\ref{hnmatrix}) could have intercell terms,
\begin{eqnarray*}
{\cal H}_{(\ell -1) p + 1 ,\ell p}^{(1,p-1)} &=&  t_{\ell L} , \\
{\cal H}_{(\ell -1) p + 2 , (\ell -1) p + 1}^{(1,p-1)} &=& t_{\ell R} , \\
{\cal H}_{(\ell - 1)p+m+2,(\ell - 1)p+m+1}^{(1,p-1)} &=& \gamma_{\ell} ,
\end{eqnarray*}
for $\ell = 1,2, \ldots , L$ and $m = 1,2,\ldots , p-2$, where $t_{\ell L}$ ($t_{\ell R}$) is the left-moving (right-moving) hopping in cell $\ell$, and $\gamma_{\ell}$ is the additional hopping in cell $\ell$.
Intracell terms could be
\begin{eqnarray*}
{\cal H}_{(\ell - 1)p+2,\ell p+1}^{(1,p-1)} &=&  J_{\ell L} , \\
{\cal H}_{\ell p+1,\ell p}^{(1,p-1)} &=& J_{\ell R} ,
\end{eqnarray*}
for $\ell = 1,2, \ldots , (L-1)$, where $J_{\ell L}$ ($J_{\ell R}$) is the left-moving (right-moving) hopping between cells $\ell$ and $\ell +1$ (all other matrix elements are zero, and we assume open boundary conditions).
For homogeneous $\gamma$ values, i.e., $\gamma_{\ell} = \gamma$ for all $\ell$, this matrix would have single-particle energy levels as Eq.~(\ref{clock}) where $\epsilon_j^{\textrm{SSH}}$, $j = 1,2,\ldots ,L$, are the positive eigenvalues of the corresponding parent SSH model.
For $t_{\ell L} \neq t_{\ell R}$ or $J_{\ell L} \neq J_{\ell R}$, the system will generally exhibit the non-Hermitian skin effect, inherited from the non-Hermitian parent model~\cite{lee16,yao18,song19,liu19b,han21,bergholtz21}. This is described in detail in the Supplemental Material~\cite{supplementary}.

Inhomogenous $\gamma$ values do not break any symmetry (including complex chiral symmetry), so the topology and spectrum of the non-Hermitian models are still closely related to that of the parent. However, the energy spectrum no longer takes the simple form of Eq.~(\ref{clock}) because there are multiple $\gamma$ values. More details are given in the Supplemental Material~\cite{supplementary}. The Supplemental Material~\cite{supplementary} also describes a study of small, random deviations away from perfect unidirectionality. In this case, the random hopping values break complex chiral symmetry. For small deviations, the bulk energy spectrum is similar to the perfect case $u=1$ described here, but the random parameter values can have a large impact on the edge states: For different realisations of disorder, their energies can lie in different positions including along either the real or the imaginary energy axis~\cite{supplementary}. Such sensitivity to parameter values is a common feature of behavior near an exceptional point~\cite{wiersig16,wiersig20}.

For simplicity, we also considered models with time-reversal symmetry~(\ref{trs}) and this constrains the complex energy spectrum to be either real or to appear with complex-conjugate pairs~\cite{kawabata19a} (it has reflection symmetry in the real energy axis).
Thus, breaking time-reversal symmetry (TRS) is likely to have an important effect for $u < 1$, depending on model details. However, it won't affect the form of the spectrum for perfect unidirectionality $u=1$, Eq.~(\ref{pdis2}), because this is determined by the complex chiral symmetry~(\ref{chiralk}).
According to the topological classification of non-Hermitian systems~\cite{kawabata19}, the topology in one dimension is the same with (class AI) and without (class A) TRS.

The Hermitian parent model has chiral symmetry so that the constructed non-Hermitian model satisfies complex chiral symmetry~(\ref{chiralk}) for $u=1$. For simplicity, we considered models with two orbitals per unit cell, but it will be possible to generalize the construction to more orbitals, and this will not effect the topological classification~\cite{kawabata19}.

\section{Conclusions}

We considered non-Hermitian Hamiltonians for non-interacting fermions with $p$ orbitals per unit cell and unidirectional hopping, generated from parent Hermitian models on a bipartite lattice. For fully unidirectional hopping, the models satisfy a complex version of chiral symmetry~(\ref{chiralk}), and the $p$ complex energy bands~(\ref{pdis2}) are given by a common $k$-dependent real factor, determined by the bands of the parent model, multiplied by the $p$th roots of unity.
When the SSH model is the parent model, the single-particle energy levels in position space~(\ref{clock}) are the same as those of free parafermion solutions to Baxter's non-Hermitian clock model~\cite{baxter89a,baxter89b,fendley14}.

For fully unidirectional hopping, it is possible to generate an arbitrary $n$th root model (with $p = 2n$ orbitals) of the parent model~\cite{arkinstall17,pelegri19,kremer20,ezawa20,lin21,marques21,viedma24,li24,huang25,zhao25,song25}.
The models support fragmented exceptional points~\cite{bid25a,bid25b}, with defective eigenvalues having an algebraic multiplicity of $p$ and a geometric multiplicity of two. When the SSH model is the parent model, defective eigenvalues are realized in position space at edges with open boundary conditions and on solitons (domain walls in the relative hopping strength).
When graphene is the parent model, defective eigenvalues occur at the corresponding Dirac points.

We described the role of partial unidirectional hopping, which breaks the complex chiral symmetry~(\ref{chiralk}).
With only real tight-binding parameters used throughout this paper, the models always satisfy time-reversal symmetry~(\ref{trs}) which ensures that the energy spectrum is either real or comes with complex-conjugate pairs~\cite{kawabata19a} (it has reflection symmetry in the real energy axis).
For an even number of orbitals $p$, the models also obey sublattice symmetry and the energy spectrum is either purely imaginary or comes with $( \epsilon , -\epsilon^{\ast} )$ pairs~\cite{kawabata19a} (it has reflection symmetry in the imaginary energy axis).

For fully bidirectional hopping, the constructed model is Hermitian, and it can be block diagonalized into even and odd parity blocks with respect to inversion of the orbitals within the unit cell.
Partially unidirectional hopping breaks the inversion symmetry and mixes the even and odd blocks, and the real energy spectrum evolves into a complex one as the degree of unidirectionality increases.
This process is determined by the topology of the parent model and by the number of orbitals per unit cell, $p$, as we described in detail for $p=3$ and $p=4$ with the example of the SSH model.
When the SSH model is the parent model, there are real energy levels due to states localized at the edges in the topological phase or on solitons.

\begin{acknowledgments}
The author thanks S.~Bid and H.~Schomerus for helpful discussions.
\end{acknowledgments}

\section*{Data availability}

The data that support the findings of this article are openly available~\cite{datanote}.

\onecolumngrid
\clearpage
\begin{center}
\textbf{\large Supplemental material: A non-Hermitian Su-Schrieffer-Heeger model with the energy levels of free parafermions}
\end{center}

\setcounter{section}{0}
\setcounter{equation}{0}
\setcounter{figure}{0}
\setcounter{table}{0}
\setcounter{page}{1}
\makeatletter
\renewcommand{\theequation}{S\arabic{equation}}
\renewcommand{\thefigure}{S\arabic{figure}}
\renewcommand{\bibnumfmt}[1]{[S#1]}
\renewcommand{\citenumfont}[1]{S#1}

\begin{center}
{\bf CONTENTS}
\end{center}
\begin{itemize}
\item[I.] Topology and partial unidirectional hopping
\begin{itemize}
\item[A.] $H^{(1,2)}$
\item[B.] $H^{(1,3)}$
\item[C.] $H^{(2,2)}$

\end{itemize}
\item[II.] Non-Hermitian skin effect
\item[III.] Inhomogeneous values of $\gamma$
\item[IV.] Small deviations from perfect unidirectionality $u=1$
\end{itemize}
\section{Topology and partial unidirectional hopping}\label{s:spartial}

Here we provide more details about the models with the SSH model as a parent, including their topology and behavior as a function of the degree of unidirectionality $u$.

\subsection{$H^{(1,2)}$}

We begin by showing that the topology at $u=1$ is determined by the parent model. The reason for this is that the relation between some of the eigenstate components in the non-Hermitian model is trivial due to the presence of the $k$-independent $\gamma$ factor.
We consider model $H^{(1,2)}$ for complete unidirectionality $u=1$,
\begin{eqnarray}
{\cal H}^{(1,2)} (k,1) = \begin{pmatrix}
0 & 0 & h^{\ast} (k) \\
h (k) & 0 & 0 \\
0 & \gamma & 0
\end{pmatrix} , \label{h12u1s}
\end{eqnarray}
We write the energy eigenvalue equation ${\cal H}^{(1,2)} \psi = \epsilon \psi$ as simultaneous equations,
\begin{eqnarray}
h^{\ast} (k) \psi_C &=& \epsilon \psi_A , \label{eq12a} \\
h (k) \psi_A &=& \epsilon \psi_B , \label{eq12b} \\
\gamma \psi_B &=& \epsilon \psi_C , \label{eq12c}
\end{eqnarray}
where the eigenstate is $\psi^T = \begin{pmatrix} \psi_A & \psi_B & \psi_C \end{pmatrix}$.
For this Hamiltonian, the energies are given by solutions of $\epsilon^3 = \gamma | h(k) |^2$. Thus, the relation between component $\psi_B$ and $\psi_C$ in Eq.~(\ref{eq12c}) is trivial as it involves a $k$-independent phase factor (only). In fact, it is possible to eliminate component $\psi_B$ using Eq.~(\ref{eq12c}) to write $\psi_B = \epsilon \psi_C / \gamma$ and substituting this into Eq.~(\ref{eq12b}). Then we can recover the Hermitian SSH model,
\begin{eqnarray}
\begin{pmatrix}
0 & h^{\ast} (k) \\
h (k) & 0
\end{pmatrix}
\begin{pmatrix}
\phi_A \\
\phi_C
\end{pmatrix}
= \varepsilon
\begin{pmatrix}
\phi_A \\
\phi_C
\end{pmatrix} ,
\end{eqnarray}
where
\begin{eqnarray}
\epsilon^3 &=& \gamma \varepsilon^2 , \\
\phi_A &=& \psi_A , \\
\phi_C &=& \sqrt{\frac{\epsilon}{\gamma}} \psi_C . 
\end{eqnarray}
Thus, topology at $u=1$ is determined by the parent model.
For $u \alt 1$, the model ${\cal H}^{(1,2)} (k,u)$ is adiabatically connected to the model for $u=1$, ${\cal H}^{(1,2)} (k,1)$, i.e., the value of $u$ can be increased to $u=1$ without breaking any relevant symmetry or closing a band gap.
However, for $u \alt 1$, the model ${\cal H}^{(1,2)} (k,u)$ is not adiabatically connected to the Hermitian model for $u=0$, ${\cal H}^{(1,2)} (k,0)$, because a band gap must close to reach $u=0$.

\begin{figure*}[t]
\includegraphics[scale=0.48]{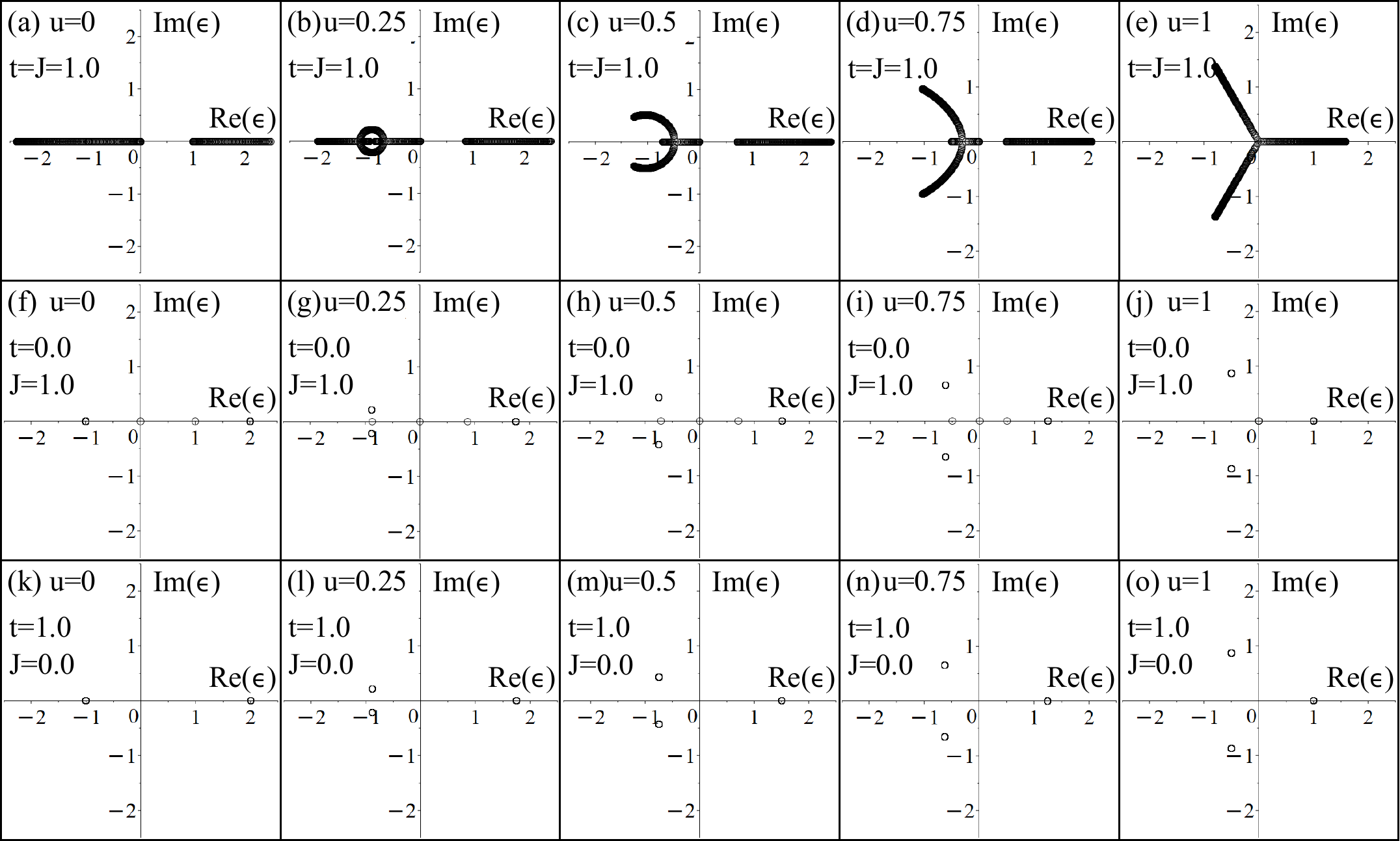}
\caption{Complex energy spectra for $p=3$ orbitals, as a function of the degree of unidirectionality $u$ for model $H^{(1,2)}$ when the SSH model is the parent model.
Energy eigenvalues (circles) are determined numerically in position space with open boundary conditions by diagonalizing the Hamiltonian. 
The top row is for $t = J = 1.0$, the middle row is for $t=0.0$, $J=1.0$, and the bottom row is for $t=1.0$, $J=0.0$. For all plots, $\gamma = 1.0$, and there are $L = 200$ unit cells.
Comparison of the middle and bottom rows shows the location of edge states with energies on the real axis:
For all $u$ values, there are three edge states with energies on the real axis (one is obscured by other energies in (f)), and they are threefold degenerate at zero energy for $u=1$ in (j).
}\label{sploth12}
\end{figure*}

This behavior is illustrated in Fig.~\ref{sploth12} where we plot the complex band spectrum for $H^{(1,2)} (k,u)$ as a function of $u$
(as in the main text Fig.~2).
The top row is for $t = J = 1.0$ so that $h(k) = 0$ for $k = \pi /a$, and the parent SSH model is in the gapless phase.
For $u=1$, the three bands meet at the origin. For $u < 1$, the band with $\mathrm{Re} (\epsilon ) > 0$ is gapped from the other two bands, but the  other two bands (with $\mathrm{Re} (\epsilon ) < 0$) are gapless.

The middle and bottom rows in Fig.~\ref{sploth12} show the complex band spectrum in the dimer limit when the band width is zero. The middle row is for the topological phase $t=0$, $J= 1.0$, and the bottom row is for the trivial phase $t=1.0$, $J=0$. The bulk bands appear to be identical in the two rows, but only the middle row shows edge states. These plots show that, in the limit of zero bandwidth, the two bands with $\mathrm{Re} (\epsilon ) < 0$ only coalesce in the $u=0$ limit (to give the Hermitian model). For all $u > 0$, the spectrum is adiabatically connected to that of $u=1$ (in the dimer limit).

\subsection{$H^{(1,3)}$}

\begin{figure*}[t]
\includegraphics[scale=0.48]{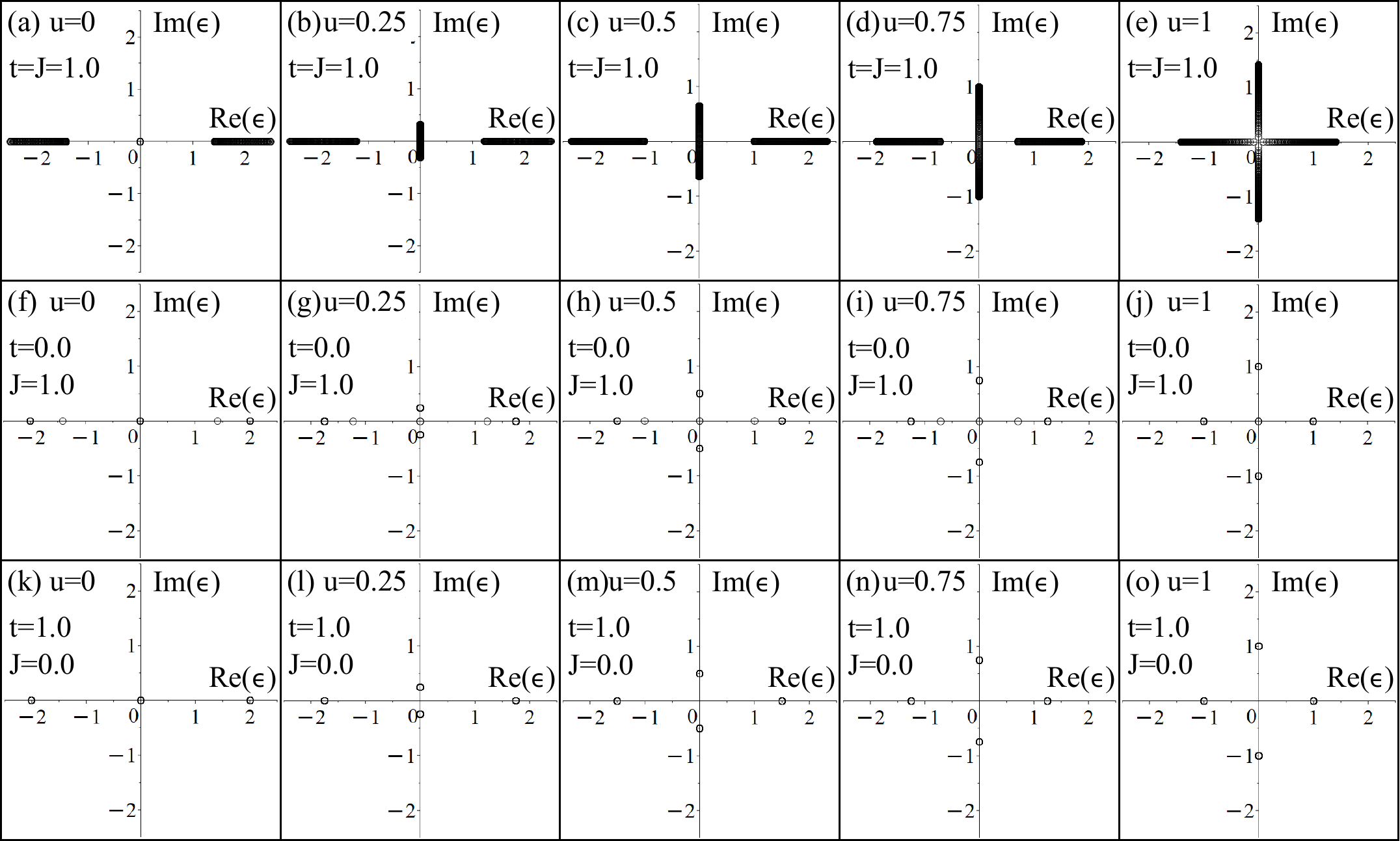}
\caption{Complex energy spectra for $p=4$ orbitals, as a function of the degree of unidirectionality $u$ for model $H^{(1,3)}$ when the SSH model is the parent model.
Energy eigenvalues (circles) are determined numerically in position space with open boundary conditions by diagonalizing the Hamiltonian. 
The top row is for $t = J = 1.0$, the middle row is for $t=0.0$, $J=1.0$, and the bottom row is for $t=1.0$, $J=0.0$. For all plots, $\gamma = 1.0$, and there are $L = 200$ unit cells.
Comparison of the middle and bottom rows shows the location of edge states with energies on the real axis: Two are degenerate at zero energy, the other two are at non-zero energy for $u < 1$. All four are degenerate at zero energy for $u=1$.
}\label{sploth13}
\end{figure*}

As in the main text Fig.~4, we plot the complex band spectrum for $H^{(1,3)} (k,u)$ as a function of $u$ in Fig.~\ref{sploth13}.
The top row is for $t = J = 1.0$ so that $h(k) = 0$ for $k = \pi /a$, and the parent SSH model is in the gapless phase.
For $u=1$, the four bands meet at the origin. For $u < 1$, the bands on the real axis are gapped from the others, but the two bands on the imaginary axis are gapless.

The middle and bottom rows in Fig.~\ref{sploth13} show the complex band spectrum in the dimer limit when the band width is zero. The middle row is for the topological phase $t=0$, $J= 1.0$, and the bottom row is for the trivial phase $t=1.0$, $J=0$. The bulk bands appear to be identical in the two rows, but only the middle row shows edge states. These plots show that, in the limit of zero bandwidth, the two bands on the imaginary axis only coalesce in the $u=0$ limit (to give the Hermitian model). For all $u > 0$, the spectrum is adiabatically connected to that of $u=1$ (in the dimer limit).

\subsection{$H^{(2,2)}$}

\begin{figure*}[t]
\includegraphics[scale=0.48]{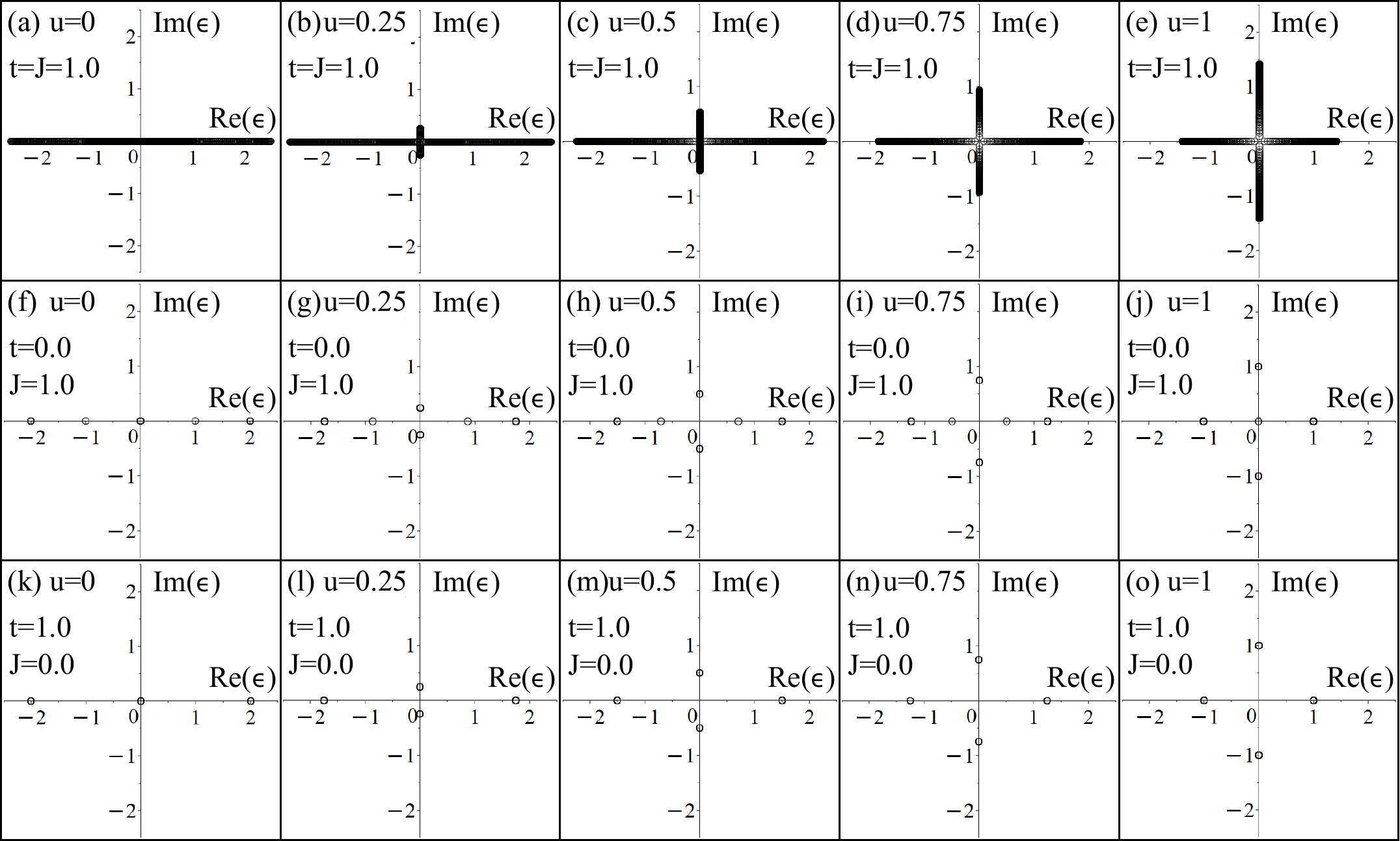}
\caption{Complex energy spectra for $p=4$ orbitals, as a function of the degree of unidirectionality $u$ for model $H^{(2,2)}$ when the SSH model is the parent model.
Energy eigenvalues (circles) are determined numerically in position space with open boundary conditions by diagonalizing the Hamiltonian. 
The top row is for $t = J = 1.0$, the middle row is for $t=0.0$, $J=1.0$, and the bottom row is for $t=1.0$, $J=0.0$. For all plots, $\gamma = 1.0$, and there are $L = 200$ unit cells.
Comparison of the middle and bottom rows shows the location of edge states with energies on the real axis:
The edge states are twofold degenerate for all $u < 1$. For $u=1$, the edge states are fourfold degenerate at zero energy.
}\label{sploth22}
\end{figure*}

As in the main text Fig.~4, we plot the complex band spectrum for $H^{(2,2)} (k,u)$ as a function of $u$ in Fig.~\ref{sploth22}.
The top row is for $t = J = 1.0$ so that $h(k) = 0$ for $k = \pi /a$, and the parent SSH model is in the gapless phase.
For $u=1$, the four bands meet at the origin. For $u < 1$, the bands remain gapless.

The middle and bottom rows in Fig.~\ref{sploth22} show the complex band spectrum in the dimer limit when the band width is zero. The middle row is for the topological phase $t=0$, $J= 1.0$, and the bottom row is for the trivial phase $t=1.0$, $J=0$. The bulk bands appear to be identical in the two rows, but only the middle row shows edge states. These plots show that, in the limit of zero bandwidth, the two bands on the imaginary axis only coalesce in the $u=0$ limit (to give the Hermitian model). For all $u > 0$, the spectrum is adiabatically connected to that of $u=1$ (in the dimer limit).

\section{Non-Hermitian skin effect}

\begin{figure}[t]
\includegraphics[scale=0.55]{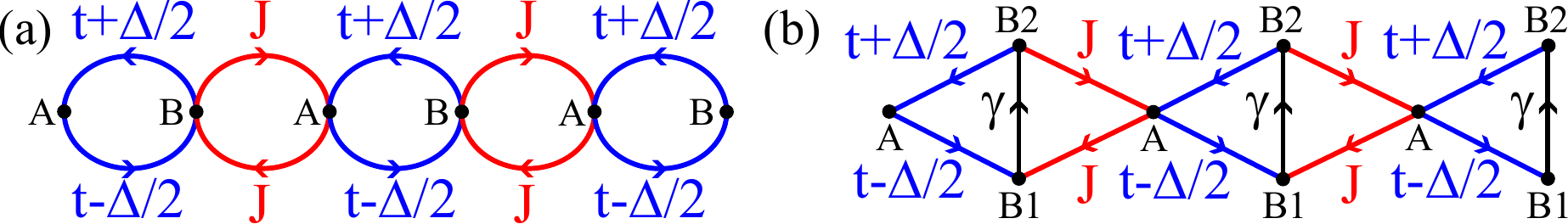}
\caption{(a) Non-Hermitian SSH model with two orbitals per unit cell on sublattices $A$ and $B$ with non-reciprocal intracell hopping $t \pm \Delta /2$ and intercell hopping $J$.
(b) Non-Hermitian Hamiltonian $H^{(1,2)}$ formed from the parent SSH model with two orbitals $B1$ and $B2$ connected by unidirectional hopping $\gamma > 0$ as indicated by the arrows, with unidirectional hopping from $A$ to $B1$ and from $B2$ to $A$, now also with the parameter $\Delta$.
}\label{smodels}
\end{figure}

As an example of the non-Hermitian skin effect~\cite{lee16s,yao18s,song19s,liu19bs,han21s,bergholtz21s}, we consider the parent SSH model to be non-Hermitian with non-reciprocal intracell hopping $t \pm \Delta / 2$ where parameter $\Delta$ characterizes the degree of non-reciprocity~\cite{yao18s}, as shown in Fig.~\ref{smodels}(a). 
In position space, the parent SSH model (Eq.~(10) in the main text) is modified as
\begin{eqnarray}
{\cal H}^{(1,1)} = \begin{pmatrix}
0 & t+\tfrac{1}{2}\Delta & 0 & 0 & 0  & \hdots & 0 & 0 & 0 \\
t-\tfrac{1}{2}\Delta & 0 & J & 0 & 0  & \hdots & 0 & 0 & 0 \\
0 & J & 0 & t+\tfrac{1}{2}\Delta & 0  & \hdots & 0 & 0 & 0 \\
0 & 0 & t-\tfrac{1}{2} & 0 & J  & \hdots & 0 & 0 & 0 \\
0 & 0 & 0 & J & 0  & \hdots & 0 & 0 & 0 \\
\vdots & \vdots & \vdots & \vdots & \vdots & \ddots & \vdots & \vdots & \vdots \\
0 & 0 & 0 & 0 & 0 & \hdots & 0 & J & 0 \\
0 & 0 & 0 & 0 & 0 & \hdots & J & 0 & t+\tfrac{1}{2}\Delta \\
0 & 0 & 0 & 0 & 0 & \hdots & 0 & t-\tfrac{1}{2} & 0
\end{pmatrix} . \label{h11s}
\end{eqnarray}

In the presence of such non-reciprocal hopping, the constructed models also display the non-Hermitian skin effect. We demonstrate this by considering model $H^{1,2}$, $p=3$, with non-reciprocal hopping as shown in Fig.~\ref{smodels}(b).
The corresponding Hamiltonian matrix, ${\cal H}^{(1,2)}$ (Eq.~(11) in the main text), is given by
\begin{eqnarray}
{\cal H}^{(1,2)} = \begin{pmatrix}
0 & 0 & t+\tfrac{1}{2}\Delta & 0 & 0 & 0 & 0 & \hdots & 0 & 0 & 0 & 0 \\
t-\tfrac{1}{2}\Delta & 0 & 0 & J & 0 & 0 & 0 & \hdots & 0 & 0 & 0 & 0 \\
0 & \gamma & 0 & 0 & 0 & 0 & 0 & \hdots & 0 & 0 & 0 & 0 \\
0 & 0 & J & 0 & 0 & t+\tfrac{1}{2}\Delta & 0 & \hdots & 0 & 0 & 0 & 0 \\
0 & 0 & 0 & t-\tfrac{1}{2}\Delta & 0 & 0 & J & \hdots & 0 & 0 & 0 & 0 \\
0 & 0 & 0 & 0 & \gamma & 0 & 0 & \hdots & 0 & 0 & 0 & 0 \\
0 & 0 & 0 & 0 & 0 & J & 0 & \hdots & 0 & 0 & 0 & 0 \\
\vdots & \vdots & \vdots & \vdots & \vdots & \vdots & \vdots & \ddots & \vdots & \vdots & \vdots & \vdots \\
0 & 0 & 0 & 0 & 0 & 0 & 0 & \hdots  & J & 0 & 0 & t+\tfrac{1}{2}\Delta \\
0 & 0 & 0 & 0 & 0 & 0 & 0 & \hdots  & 0 & t-\tfrac{1}{2}\Delta & 0 & 0 \\
0 & 0 & 0 & 0 & 0 & 0 & 0 & \hdots  & 0 & 0 & \gamma & 0
\end{pmatrix} . \label{hp3s}
\end{eqnarray}

\begin{figure}[t]
\includegraphics[scale=0.20]{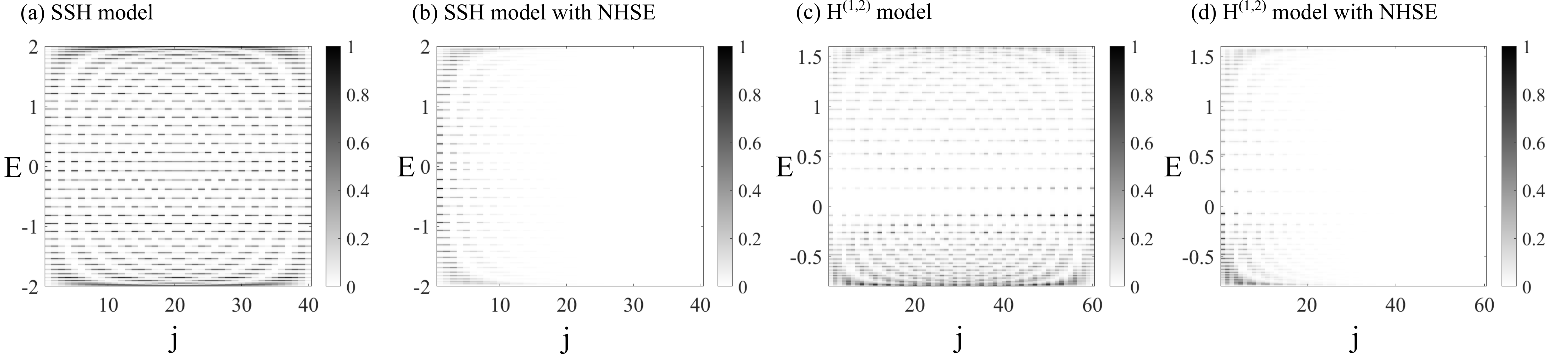}
\caption{Plots of the local density of states projected onto the real energy axis~(\ref{localdos}), $g ( j , E)$, for real energy $E$ and lattice site with index $j$ illustrating the non-Hermitian skin effect (NHSE).
(a) is the Hermitian SSH model ($\Delta = 0$), (b) is a non-Hermitian SSH model with non-reciprocal hopping  ($\Delta = 0.5$), Fig.~\ref{smodels}(a), (c) is model $H^{(1,2)}$ ($\Delta = 0$), (d) is model $H^{(1,2)}$ with additional non-reciprocal hopping ($\Delta = 0.5$), Fig.~\ref{smodels}(b).
Energy eigenvalues and right eigenstates are determined numerically in position space with open boundary conditions by diagonalizing the Hamiltonian, Eq.~(\ref{h11s}) for (a) and (b), Eq.~(\ref{hp3s}) for (c) and (d). 
For (a) and (b), parameter values are $t = J = 1.0$, and there are $20$ unit cells corresponding to $40$ lattice sites.
For (c) and (d), parameter values are $t = J = \gamma = u = 1.0$, and there are $20$ unit cells corresponding to $60$ lattice sites.
To determine the local density of states, we consider $1001$ $E$ values with Lorentzian width $\delta = 0.005$.
}\label{snonrecipfig}
\end{figure}

To illustrate the non-Hermitian skin effect, we numerically diagonalize the Hamiltonians and determine the local density of states projected onto the real energy axis, $g ( j , E)$, for real energy $E$ and lattice site with index $j$,
\begin{eqnarray}
g ( j , E) \approx \frac{\delta}{\pi} \sum_n \frac{|\psi_{j,n}|^2}{(E - \mathrm{Re} (\epsilon_n))^2 + \delta^2} , \label{localdos}
\end{eqnarray}
where we approximate a Dirac delta function with a Lorentzian of width $\delta$. We consider the real part of an eigenenergy $\epsilon_n$ with eigenstate component $\psi_{j,n}$ on lattice site $j$ for a finite system in position space with open boundary conditions.
This is the right eigenstate, i.e., the solution of ${\cal H} \psi_{j,n} = \epsilon_n \psi_{j,n}$.
Representative numerical data is plotted in Fig.~\ref{snonrecipfig}.
In (a) and (b), we consider the SSH model with and without reciprocal hopping, respectively. Fig.~\ref{snonrecipfig}(a) shows bulk eigenstates extended across the whole sample whereas Fig.~\ref{snonrecipfig}(b) illustrates the non-Hermitian skin effect with all eigenstates localized towards the left edge of the system.
Similar behavior occurs for the constructed non-Hermitian model $H^{(1,2)}$ shown in Fig.~\ref{snonrecipfig}(c) and (d) with and without additional reciprocal hopping, respectively. Fig.~\ref{snonrecipfig}(c) shows bulk eigenstates extended across the whole sample whereas Fig.~\ref{snonrecipfig}(d) illustrates the non-Hermitian skin effect with all eigenstates localized towards the left edge of the system.

\section{Inhomogeneous values of $\gamma$}

Inhomogenous $\gamma$ values do not break any symmetry (including complex chiral symmetry), so the topology and spectrum of the non-Hermitian models are still closely related to that of the parent. However, the energy spectrum no longer takes the simple form of Eq.~(12) in the main text because there are multiple $\gamma$ values.

\begin{figure}[t]
\includegraphics[scale=0.50]{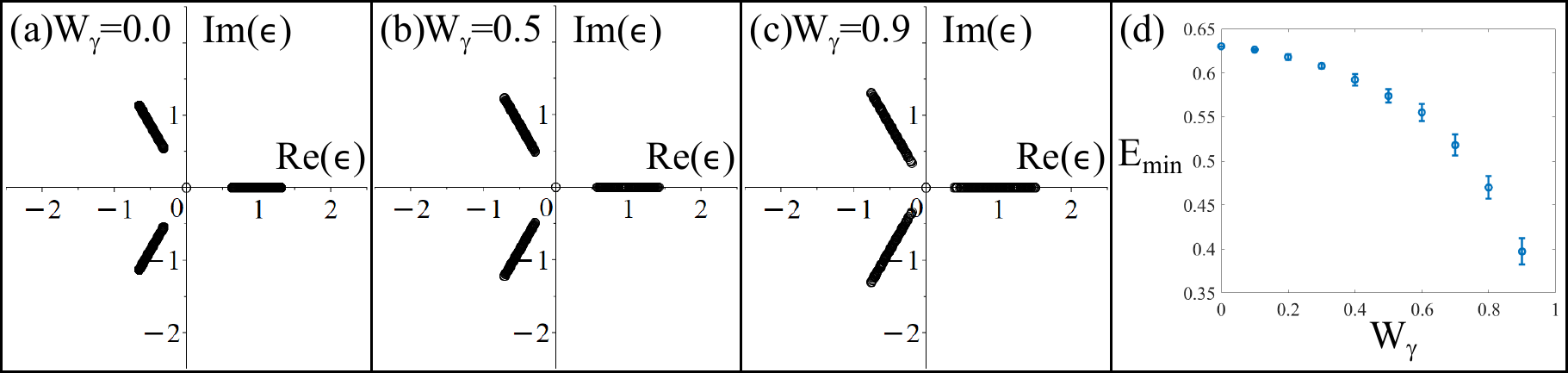}
\caption{Model $H^{(1,2)}$ in the presence of inhomogeneous values of $\gamma$. For the $j$th unit cell, the value of $\gamma$ is $\gamma_j = 1.0 + W_{j}$ where $W_{j}$ is a random value uniformly distributed in the range $[-W_{\gamma} , W_{\gamma} ]$.
(a), (b) and (c) show the complex energy spectra, each for a single typical realization of disorder with values $W_{\gamma} = 0.0$, $0.5$, $0.9$, respectively.
(d) shows the the energy of the smallest positive real bulk eigenvalue $E_{\mathrm{min}}$ as a function of disorder strength $W_{\gamma}$, averaged over $20$ disorder realizations.
In all plots, energy eigenvalues are determined numerically in position space with open boundary conditions by diagonalizing the Hamiltonian. 
Parameter values are $t=0.5$, $u = J = 1.0$, and there are $L = 200$ unit cells.
}\label{sgammaplot}
\end{figure}

To illustrate these points, Fig.~\ref{sgammaplot} shows model $H^{(1,2)}$ in the presence of random values of $\gamma$. For the $j$th unit cell, the value of $\gamma$ is $\gamma_j = 1.0 + W_{j}$ where $W_{j}$ is a random value uniformly distributed in the range $[-W_{\gamma} , W_{\gamma} ]$.
Fig.~\ref{sgammaplot}(a), (b) and (c) show the complex energy spectra for a single typical realization of disorder with values $W_{\gamma} = 0.0$, $0.5$, $0.9$, respectively. This shows that the spectra still obey complex chiral symmetry, but the bandwidth generally increases with the disorder strength $W_{\gamma}$. To show this, Fig.~\ref{sgammaplot}(d) plots the value of the smallest positive real bulk eigenvalue $E_{\mathrm{min}}$ as a function of disorder strength $W_{\gamma}$, averaged over $20$ disorder realizations.  This energy decreases with disorder, consistent with an increase in bandwidth.

\section{Small deviations from perfect unidirectionality $u=1$}

We study the effect of small deviations from perfect unidirectionality by considering ${\cal H}^{(1,2)} (u=1)$ plus its Hermitian conjugate, where the Hermitian conjugate part contains random hopping parameters. The corresponding Hamiltonian matrix, ${\cal H}^{(1,2)}$ (Eq.~(11) in the main text), is given by
\begin{eqnarray}
{\cal H}^{(1,2)} = \begin{pmatrix}
0 & w_{t1}t & t & 0 & 0 & 0 & 0 & \hdots & 0 & 0 & 0 & 0 \\
t & 0 & w_{\gamma 1}\gamma & J & 0 & 0 & 0 & \hdots & 0 & 0 & 0 & 0 \\
w_{t1}t & \gamma & 0 & w_{J1}J & 0 & 0 & 0 & \hdots & 0 & 0 & 0 & 0 \\
0 & w_{J1}J & J & 0 & w_{t2}t & t & 0 & \hdots & 0 & 0 & 0 & 0 \\
0 & 0 & 0 & t & 0 & w_{\gamma 2}\gamma & J & \hdots & 0 & 0 & 0 & 0 \\
0 & 0 & 0 & w_{t2}t & \gamma & 0 & w_{J2}J & \hdots & 0 & 0 & 0 & 0 \\
0 & 0 & 0 & 0 & w_{J2}J & J & 0 & \hdots & 0 & 0 & 0 & 0 \\
\vdots & \vdots & \vdots & \vdots & \vdots & \vdots & \vdots & \ddots & \vdots & \vdots & \vdots & \vdots \\
0 & 0 & 0 & 0 & 0 & 0 & 0 & \hdots  & J & 0 & w_{tL}t & t \\
0 & 0 & 0 & 0 & 0 & 0 & 0 & \hdots  & 0 & t & 0 & w_{\gamma L}\gamma \\
0 & 0 & 0 & 0 & 0 & 0 & 0 & \hdots  & 0 & w_{tL}t & \gamma & 0
\end{pmatrix} . \label{hus}
\end{eqnarray}
This means that, for the Hermitian conjugate part, intracell hopping in cell $\ell$ is $t_{\ell} = w_{t\ell}t$ where $w_{t\ell}$ is a random value uniformly distributed in the range $[-W_u , W_u ]$, auxiliary hopping is $\gamma_{\ell} = w_{\gamma\ell}\gamma$ where $w_{\gamma\ell}$ is a random value uniformly distributed in the range $[-W_u , W_u ]$, and intercell hopping from cell $\ell$ to $\ell + 1$ is $\gamma_{\ell} = w_{\gamma\ell}\gamma$ where $w_{\gamma\ell}$ is a random value uniformly distributed in the range $[-W_u , W_u ]$.
For non-zero disorder strength $W_u$, this Hamiltonian breaks complex chiral symmetry, but it satisfies time-reversal symmetry and pseudo-hermicity for $p=3$ (for even $p$, the corresponding Hamiltonians would satisfy sublattice symmetry).

\begin{figure}[t]
\includegraphics[scale=0.50]{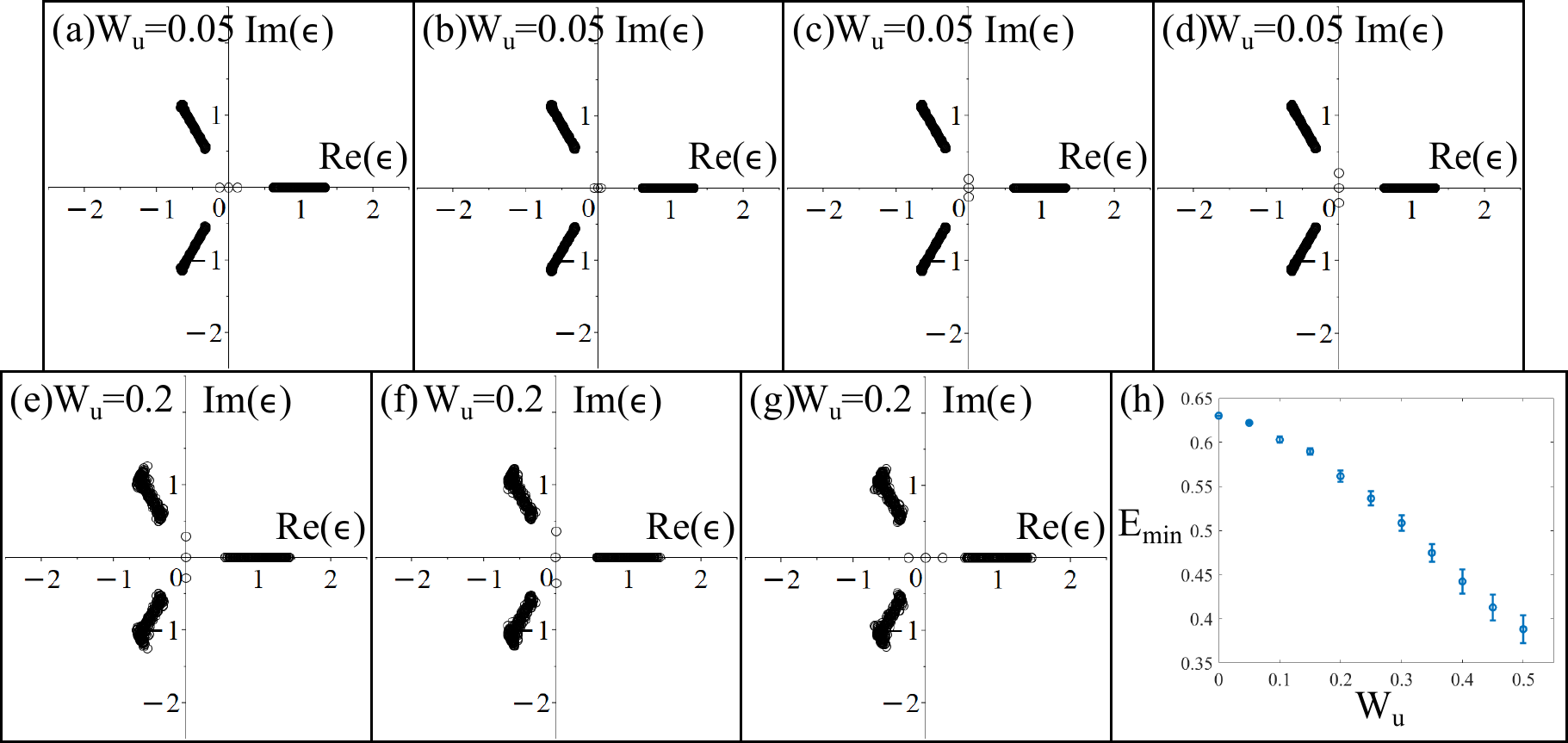}
\caption{Model $H^{(1,2)}$ in the presence of random deviations from perfect unidirectionality as described by the Hamiltonian~(\ref{hus}).
(a)-(d) show the complex energy spectra, each for a single typical realization of disorder with the same disorder value $W_u = 0.05$,
(e), (f) and (g) show the complex energy spectra, each for a single typical realization of disorder with the same disorder value $W_u = 0.2$,
(h) shows the the energy of the smallest positive real bulk eigenvalue $E_{\mathrm{min}}$ as a function of disorder strength $W_u$, averaged over $20$ disorder realizations.
In all plots, energy eigenvalues are determined numerically in position space with open boundary conditions by diagonalizing the Hamiltonian~(\ref{hus}). 
Parameter values are $t=0.5$, $\gamma = J = 1.0$, and there are $L = 200$ unit cells.
}\label{suplot}
\end{figure}

Fig.~\ref{suplot} shows model $H^{(1,2)}$ in the presence of non-zero disorder strength $W_u$.
Fig.~\ref{suplot}(a)-(d) show the complex energy spectra, each for a single typical realization of disorder with the same disorder value $W_u = 0.05$.
Fig.~\ref{suplot}(e), (f) and (g) show the complex energy spectra, each for a single typical realization of disorder with the same disorder value $W_u = 0.2$. These plots illustrate that complex chiral symmetry is broken because the three bands do not lie exactly along the lines with arguments $0$ and $\pm 2 \pi i / 3$. Although the bulk bands of the plots look roughly similar, the positions of the levels corresponding to edge states (the three isolated circles near the origin) can be in different positions including being oriented along either the imaginary or the real axis, depending on the disorder realization, even at the low disorder value $W_u = 0.05$. Such sensitivity to parameter values is common behavior near exceptional points~\cite{wiersig16s,wiersig20s}. Note that the spectra all have reflection symmetry with respect to the real energy axis, in accordance with time-reversal symmetry. Fig.~\ref{suplot}(h) plots the value of the smallest positive real bulk eigenvalue $E_{\mathrm{min}}$ as a function of disorder strength $W_u$, averaged over $20$ disorder realizations, showing that $E_{\mathrm{min}}$ decreases with disorder.

\end{document}